\documentclass[a4paper,DIV=12]{scrartcl}

\setkomafont{disposition}{\normalfont\bfseries}

\usepackage{amsmath}
\usepackage{amsthm}
\usepackage{amssymb}
\usepackage{stmaryrd}
\usepackage[ruled,vlined]{algorithm2e}
\usepackage{mathrsfs}
\usepackage{enumerate}
\usepackage{mathtools}
\usepackage{xspace}
\usepackage{float}
\usepackage{url}
\usepackage{todonotes}
\usepackage{multicol}
\usepackage{listofitems}
\usepackage{tikz-cd}
\usepackage[font=sf,width=.9\linewidth]{caption}
\usepackage[labelformat=simple]{subcaption}

\title{Conelikes and Ranker Comparisons}

\author{Viktor Henriksson\hspace*{1pt}$^1$ and Manfred Kuf\-leitner\hspace*{2pt}$^2$}
\date{\normalsize$^1$ Loughborough University, Loughborough, UK \\
\texttt{b.v.d.henriksson@lboro.ac.uk} \\[3mm]
$^2$ University of Stuttgart, Stuttgart, Germany \\
\texttt{kufleitner@fmi.uni-stuttgart.de}}

\newtheorem{proposition}{Proposition}
\newtheorem{theorem}{Theorem}
\newtheorem{corollary}{Corollary}
\newtheorem{lemma}{Lemma}
\theoremstyle{definition}
\newtheorem{definition}{Definition}
\newtheorem{example}{Example}

\newcommand{\FO}{\textnormal{FO}}

\newcommand{\DA}{\mathbf{DA}}
\newcommand{\psat}[2]{\mathsf{ConeSat}_{#1}(#2)}
\newcommand{\satFO}[2]{\mathsf{Sat}_{\mathbf{R}_{#1} \cap \mathbf{L}_{#1}}(#2)}
\newcommand{\satSP}[2]{\mathsf{Sat}_{\mathbf{R}_{#1} \vee \mathbf{L}_{#1}}(#2)}
\newcommand{\psatS}[2]{\mathsf{ConeSat}_{\mathbf{Si}_{#1}}(#2)}
\newcommand{\psatP}[2]{\mathsf{ConeSat}_{\mathbf{Pi}_{#1}}(#2)}
\newcommand{\cone}[2]{\mathsf{Cone}_{#1}(#2)}
\newcommand{\coneS}[2]{\mathsf{Cone}_{\mathbf{Si}_{#1}}(#2)}
\newcommand{\coneP}[2]{\mathsf{Cone}_{\mathbf{Pi}_{#1}}(#2)}

\newcommand{\plFO}[2]{\mathsf{PL}_{\mathbf{R}_{#1 + 1} \cap \mathbf{L}_{#1 + 1}}(#2)}
\newcommand{\plSP}[2]{\mathsf{PL}_{\mathbf{R}_{#1} \vee \mathbf{L}_{#1}}(#2)}
\newcommand{\satR}[2]{\mathsf{Sat}_{\mathbf{R}_{#1}}(#2)}
\newcommand{\plR}[2]{\mathsf{PL}_{\mathbf{R}_{#1}}(#2)}
\newcommand{\satL}[2]{\mathsf{Sat}_{\mathbf{L}_{#1}}(#2)}
\newcommand{\plL}[2]{\mathsf{PL}_{\mathbf{L}_{#1}}(#2)}
\newcommand{\sat}[2]{\mathsf{Sat}_{#1}(#2)}
\newcommand{\pl}[2]{\mathsf{PL}_{#1}(#2)}
\newcommand{\relmorsmall}[6]{#1:#3 \xleftarrow{#5} #2 \xrightarrow{#6} #4}
\newcommand{\alp}{\mathsf{alph}}
\newcommand{\faktor}[2]{#1 / \!\! #2}
\newcommand{\intp}[1]{\llbracket #1 \rrbracket}

\newcommand{\ceil}[1]{\lceil #1 \rceil}
\renewcommand{\th}{\ensuremath{^{\text{th}}}}

\newcommand{\Si}{\mathbf{Si}}
\newcommand{\Pivar}{\mathbf{Pi}}

\newcommand{\greenfont}[1] {\ensuremath{\mathcal{#1}}}

\newcommand{\greenR} {\greenfont{R}\xspace}
\newcommand{\Req}    {\mathrel{\greenR}}
\newcommand{\Rleq}   {\mathrel{\leq_\greenR}}
\newcommand{\Rl}     {\mathrel{<_\greenR}}

\newcommand{\greenL} {\greenfont{L}\xspace}
\newcommand{\Leq}    {\mathrel{\greenL}}
\newcommand{\Lleq}   {\mathrel{\leq_\greenL}}

\newcommand{\greenJ} {\greenfont{J}\xspace}
\newcommand{\Jeq}    {\mathrel{\greenJ}}
\newcommand{\Jleq}   {\mathrel{\leq_\greenJ}}
\newcommand{\Jl}     {\mathrel{<_\greenJ}}

\newcommand{\RR}{\textnormal{R\!R}}
\newcommand{\XX}{\textnormal{X\!X}}
\newcommand{\YY}{\textnormal{Y\!Y}}
\newcommand{\XY}{\textnormal{X\!Y}}
\newcommand{\YX}{\textnormal{Y\!X}}

\newcommand{\itref}[1]{\textit{(\ref{#1})}}

\newcounter{resumeCounter}

\begin{document}

\maketitle

\begin{abstract}
For every fixed class of regular languages, there is a natural hierarchy of increasingly more general problems: Firstly, the membership problem asks whether a given language belongs to the fixed class of languages. Secondly, the separation problem asks for two given languages whether they can be separated by a language from the fixed class. And thirdly, the covering problem is a generalization of separation problem to more than two given languages. Most instances of such problems were solved by the connection of regular languages and finite monoids. Both the membership problem and the separation problem were also extended to ordered monoids. The computation of pointlikes can be interpreted as the algebraic counterpart of the separation problem. In this paper, we consider the extension of computation of pointlikes to ordered monoids. This leads to the notion of conelikes for the corresponding algebraic framework. 

We apply this framework to the Trotter-Weil hierarchy and both the full and the half levels of the $\FO^2$ quantifier alternation hierarchy. As a consequence, we solve the covering problem for the resulting subvarieties of $\DA$. An important combinatorial tool are uniform ranker characterizations for all subvarieties under consideration; these characterizations stem from order comparisons of ranker positions.
\end{abstract}

\section{Introduction}

For a given variety of regular languages, there is a hierarchy of decision problems: First, we can ask whether a given regular language is in the variety; this is known as the \emph{membership problem}. Very often, the membership problem is solved by giving an effective characterization. Famous solutions to the membership problem includes Simon's characterization of the piecewise testable languages in terms of \greenJ-trivial monoids~\cite{simon1975lncs}, and Sch\"utzenberger's characterization of the star-free languages by aperiodic monoids~\cite{shutzenberger1976sf}. Inspired by these results, Eilenberg showed that there exists a one to one correspondence between varieties of regular languages and varieties of finite monoids \cite{eilenberg1974}. This correspondence leads to an important approach for deciding the membership problem: one verifies some equivalent algebraic property of the syntactic monoid. The challenge here, however, is to identify the algebraic property and to prove its equivalence.

A more general problem is the \emph{separation problem}. Given two languages, it asks whether there exists a language in the fixed variety which contains the first language and is disjoint with the second language. By applying the separation problem to a language and its complement, we obtain an answer to the membership problem. Thus, the separation problem is more general than the membership problem. Moreover, the separation problem can be used as a tool to solve the membership problem for varieties where this was not previously known; see e.g.~\cite{PlaceZeitoun2014icalp}.
A further generalization is given by the \emph{covering problem} \cite{PlaceZeitoun2018lmcs}. This problem considers a finite set of languages and a distinguished language, and asks how well the finite set of languages can be separated by a cover of the distinguished language.

As noted by Almeida, the separation problem for regular languages can also be solved via algebra by deciding so-called pointlikes~\cite{Almeida1999pms}. 
The problem of deciding pointlikes is well studied, and there are effective characterizations for many varieties, e.g.\ aperiodics \cite{Henckell1988jpaa}, \greenR-trivial monoids \cite{AlmeidaSilva2001tcs}, \greenJ-trivial monoids \cite{AlmeidaZeitoun1997rairo,Steinberg1998ijac} and finite groups \cite{Ash1991ijac}.

A well studied fragment of first order logic is two-variable first-order logic $\FO^2$. The languages definable in $\FO^2$ form a variety, with the corresponding monoid variety $\DA$. In the study of $\FO^2$ and $\DA$, two natural hierarchies have emerged: the Trotter-Weil hierarchy defined by a deep connection to the hierarchy of bands, and the quantifier alternation hierarchy. In stark contrast to the full $\FO$ quantifier alternation hierarchy, membership of the $\FO^2$ quantifier alternation hierarchy is solved for all levels \cite{kufleitnerweil2012csl,KrebsStraubing2017tocl,fleischerkufleitnerlauser2014csr}. In particular, a tight connection between the Trotter-Weil and the quantifier alternation hierarchy has appeared; Weil and the second author showed that the join levels of the quantifier alternation hierarchy (i.e., the $\FO^2_m$ levels) correspond to the intersection levels of the Trotter-Weil hierarchy \cite{kufleitnerweil2012csl}, and combining two results from~\cite{KufleitnerLauser2012lncs} and~\cite{fleischerkufleitnerlauser2014csr} shows that the join levels of the Trotter-Weil hierarchy correspond to the intersection levels of the quantifier alternation hierarchy.

Rankers have emerged as an important tool in the study of $\FO^2$. These were first introduced by Schwentick, Thérien, and Vollmer by the name of turtle programs \cite{SchwentickTherienVollmer2001dlt}. Using comparisons of restricted sets of rankers, Weis and Immerman gave a combinatorial characterization of the full levels of the quantifier alternation hierarchy \cite{WeisImmerman2009lmcs}. This approach was extended to the half-levels of the quantifier alternation hierarchy in the PhD-thesis of Lauser~\cite{Lauser2014thesis}. The corners of the Trotter-Weil hierarchy also admit ranker characterizations using the concept of condensed rankers \cite{KufleitnerWeil2012lmcs}.

This article solves the covering problem (and thus also separation problem) for all levels of the Trotter-Weil hierarchy and quantifier alternation hierarchy inside $\FO^2$. For this, we rely on two main tools, \emph{conelikes} and \emph{ranker comparisons}. 
Conelikes are introduced in Section \ref{sec:Conelikes}. They extend pointlikes to ordered monoids, and are algebraic versions of the imprints used by Place and Zeitoun; see e.g.~\cite{PlaceZeitoun2018lmcs}. Thus, they have a strong connection to the covering problem; an algorithm for computing the conelikes with respect to a monoid variety can be used to solve the covering problem for the corresponding language variety and vice versa.

Sections \ref{sec:rankercomp} and \ref{sec:RankerCompHierarchy} deals with ranker comparisons. In Section \ref{sec:rankercomp}, we give a framework for ranker comparisons using general sets of ranker pairs. We show that any set of pairs of rankers which is closed under ranker subwords gives rise to a stable relation and thus defines a monoid.

In Section \ref{sec:RankerCompHierarchy}, we use this framework to give uniform characterizations for all levels of the Trotter-Weil and quantifier alternation hierarchy. In particular, we give a characterization of the corners of the Trotter-Weil hierarchy in terms of ranker comparisons.
Together, these sets of ranker comparisons form a natural hierarchy, the \emph{ranker comparison hierarchy} which encompasses both the quantifier alternation hierarchy and the Trotter-Weil hierarchy.

The rest of the article is devoted to the solution of the covering problem. In Section \ref{sec:decidingconelikes}, we present sets of subsets of a monoid which can be computed effectively. Our main theorem states that these sets coincide with the conelikes (or the pointlikes for the unordered varieties). Our main theorem also provides optimal separators: relational morphisms such that the conelikes with respect to these morphisms are the same as the conelikes with respect to the corresponding variety. The co-domains in these morphisms are defined using ranker comparisons.

After handling two special cases in Section \ref{sec:BaseCases}, we use Section \ref{sec:EasyDirection} to show that being in the introduced sets implies being conelike with respect to the relevant variety.
The more difficult direction is showing that every conelike lies in the introduced sets. In Section \ref{sec:OnStructure}, we show how languages defined by ranker comparisons can be factored in a convenient way, and in Section \ref{sec:DifficultDirection}, we use these factorizations to complete the circle. We show that given a monoid defined by the proper set of ranker comparisons, the conelikes with respect to this monoid are all in the corresponding set. In particular, this means that every conelike with respect to the corresponding monoid variety is in the set.

\section{Preliminaries}

\subsection{Words and Monoids}
\label{sec:wordsandmonoids}

If $M$ is a semigroup and $e \in M$ satisfies $ee = e$, then $e$ is \emph{idempotent}. Given a semigroup $M$ there exists a (smallest) number $\omega_M$ such that $u^{\omega_M}$ is idempotent for each $u \in M$. If $M$ is clear from context, we write $\omega$ for this number.
For sets $S, T \subseteq M$, we have
\begin{equation*}
	ST = \left\{ st \in M \mid s \in S, t \in T \right\}.
\end{equation*}
Note that $2^M$ is a monoid under this operation.

An important tool in semigroup theory are the \emph{Green's relations}, out of which we introduce the following three. Given a monoid $M$ and $s,t \in M$, we define
\begin{itemize}
	\item $s \Rleq t$ if $sM \subseteq tM$,
	\item $s \Lleq t$ if $Ms \subseteq Mt$,
	\item $s \Jleq t$ if $MsM \subseteq MtM$.
\end{itemize}
We define $s \Req t$ if $s \Rleq t$ and $t \Rleq s$ and we define $s \Leq t$ and $s \Jeq t$ correspondingly. We say that $s \Rl t$ if $s \Rleq t$ but not $s \Req t$ and equivalently for \greenL and \greenJ.  
Let $u \in A^*$ and $\mu: A^* \to M$. Then there is a unique factorization $u = u_1 a_1 \dots u_{n-1} a_{n-1} u_n$ such that $\mu(u_1 a_1 \dots a_i) \Req \mu(u_1 a_1 \dots a_i u_{i+1}) \Rl \mu(u_1 a_1 \dots a_i u_{i+1} a_{i+1})$. This is the \emph{\greenR-factorization of $u$ with respect to $\mu$}. The \emph{\greenL-factorization of $u$ with respect to $\mu$} is defined symmetrically.

Given a monoid $M$ with a binary relation $\preceq$, we say that $\preceq$ is \emph{stable} if for all $s,t,x,y \in M$, $s \preceq t$ implies $xsy \preceq xty$. We say that a monoid is \emph{ordered} if it is equipped with a stable order. A \emph{congruence} is a stable equivalence relation. In particular, any stable preorder $\preceq$ induces the congruence given by $s \sim t$ if and only if $s \preceq t$ and $t \preceq s$. If $M$ is ordered, and $s \in M$, then $\uparrow s = \left\{ t \in M \mid s \leq t \right\}$.
If $M$ is a monoid, and $\preceq$ is a stable preorder, then $\faktor{M}{\preceq}$ is the monoid whose elements are the equivalence classes of the induced congruence, the multiplication is that induced by the multiplication in $M$ and where, for $s,t \in M$ with $[s],[t]$ the corresponding equivalence classes, we have $[s] \leq [t]$ if and only if $s \preceq t$.
Given a language $L$, the \emph{syntactic preorder} is the relation $u \leq_{L} v$ if and only if $xuy \in L \Rightarrow xvy \in L$ for all $x,y \in A^*$. Let $\mu: A^* \to \faktor{A^*}{\leq_L}$ be the canonical projection, then $\pi$ is the \emph{syntactic morphism} and $\faktor{A^*}{\leq_L}$ the \emph{syntactic monoid} of $L$. A language is \emph{regular} if and only if the syntactic monoid is finite.

Let $M$ and $N$ be (possibly ordered) monoids. A \emph{relational morphism} is a relation $\tau: M \to N$ (or mapping $M \to 2^N$) which satisfies
\begin{enumerate}[(i)]
		\item $1_N \in \tau(1_M)$,
		\item for all $s \in M$, $\tau(s) \neq \emptyset$,
		\item for all $s,t \in M$, $\tau(s)\tau(t) \subseteq \tau(st)$.
	\end{enumerate}
If there is a relational morphism $\tau: M \to N$ such that $\tau(s) \cap \tau(s') \neq \emptyset$ implies $s = s'$ we say that $M$ \emph{divides} $N$. A \emph{division of ordered monoids} is a division where we also assume $t \leq t'$ for some $t \in \tau(s)$, $t' \in \tau(s')$ implies $s \leq s'$. A \emph{variety of monoids} is a collection of monoids closed under division and finite direct products. A collection of ordered monoids is a \emph{positive variety} if it is closed under finite direct products and division of ordered monoids.

For a relational morphism $\tau: M \to N$, a set $S \subseteq M$ such that $\bigcap_{s \in S} \tau(s) \neq \emptyset$ is \emph{pointlike} with respect to $\tau$. If $t \in \bigcap_{s \in S} \tau(s)$, then $t$ is a \emph{witness} of $S$ being pointlike. If $\mathbf{V}$ is a variety and $S$ is pointlike for every relational morphism $\tau: M \to N \in \mathbf{V}$, then $S$ is pointlike with respect to $\mathbf{V}$. The set of all pointlikes in $M$ with respect to $\tau$ is $\pl{\tau}{M}$, and the set of all pointlikes with respect to $\mathbf{V}$ is $\pl{\mathbf{V}}{M}$.

A useful way to define varieties is through the use of \emph{$\omega$-identities} and \emph{$\omega$-relations}. An \emph{$\omega$-term} is either $x$ where $x$ is taken from some (usually infinite) set of variables $X$, or $tt'$ or $t^{\omega}$ where $t$ and $t'$ are $\omega$-terms. An $\omega$-identity is given by $t = t'$ or $t \leq t'$ where $t$ and $t'$ are $\omega$-terms. Given a monoid $M$, an \emph{interpretation} of $\omega$-terms is any extension of a map $\chi: X \to M$ for which $\chi(tt') = \chi(t)\chi(t')$ and $\chi(t^{\omega}) = \chi(t)^{\omega_M}$. We say that a monoid $M$ \emph{satisfy} an $\omega$-identity $t = t'$ if $\chi(t) = \chi(t')$ for all interpretations $\chi$. It similarly satisfies $t \leq t'$ if $\chi(t) \leq \chi(t')$ for all interpretations.
If $R_1,\dots,R_n$ are $\omega$-identities or -relations, then $\intp{R_1,\dots,R_n}$ denotes the collection of all monoids which satisfy all $R_i$.
Some varieties that are of importance in this text are
\begin{itemize}
	\item $\DA = \intp{(xzy)^{\omega} = (xzy)^{\omega}z(xzy)^{\omega}}$.
	\item $\mathbf{J} = \intp{(st)^{\omega}s(xy)^{\omega} = (st)^{\omega}y(xy)^{\omega}}$, or equivalently all monoids whose \greenJ-classes are trivial.
	\item $\mathbf{J}_1 = \intp{x^2 = x, xy = yx}$,
	\item $\mathbf{J}^+ = \intp{1 \leq z}$.
\end{itemize}

	Let $A$ be a collection of symbols, called an \emph{alphabet}. We denote by $A^*$ the set of concatenations of symbols in $A$. In other words, $A^*$ is the \emph{free monoid} of $A$. An element $u \in A^*$ is a \emph{word} and a subset $L \subseteq A^*$ a \emph{language}.
We will denote the empty word by $\varepsilon$.
A \emph{(scattered) subword} of $u$ is a word $v = a_1 \dots a_n$ such that $u = u_1 a_1 \dots u_n a_n u_{n+1}$ for some (possibly empty) words $u_{i}$. Let $u = u_1 u_2 u_3$ for some (possibly empty) words $u_1$, $u_2$, $u_3$. Then $u_1$ is a \emph{prefix} and $u_2$ is a \emph{factor} of $u$.

If $u = u_1 \dots u_n$ where each $u_i$ is a word, then $u_1 \dots u_n$ is a \emph{factorization} of $u$. This concept extends to subsets of $A^*$; if $U \subseteq A^*$, a factorization of $U$ is $U_1 \dots U_n$ where each $u \in U$ can be factored as $u = u_1 \dots u_n$ in such a way that $u_i \in U_i$. This definition does not coincide with the monoid operation on subsets given above.\footnote{Indeed, $\left\{ aa,bb \right\}$ can be factored as $\left\{ a,b \right\}\left\{ a,b \right\}$, but $ab, ba \in \left\{ a,b \right\}\left\{ a,b \right\}$ if seen as a multiplication.} To resolve this ambiguity, we always consider concatenation to mean factorization when dealing with $A^*$, unless otherwise specified.

For an alphabet $A$, let $J_A$ denote the monoid whose elements are subsets of $A$ and whose operation is the union operation. This monoid has a natural ordering defined by $U \leq V$ if $U \subseteq V$ for $U, V \subseteq A$. Let $\alp_A: A^* \to J_A$ be the extension of $\alp_A(a) = \left\{ a \right\}$ for each $a \in A$. We will drop the subscript when $A$ is clear from context. Given a surjective homomorphism $\mu: A^* \to M$, a morphism $\alpha: M \to J_A$ is called a \emph{content morphism} if $\alp_A = \alpha \circ \mu$. 

	Let $\tau$ be a relation, and let $G = \left\{ (s,t) \in M \times N \mid t \in \tau(s) \right\}$ be its graph. Then $\tau$ is a relational morphism if and only if $G$ is a submonoid of $M \times N$ and the projection on $M$ is surjective.
	Any relational morphism can be thought of as pulling the elements of $M$ back to some free monoid $A^*$ where $A$ generates both $M$ and $N$, and then pushing the words to their corresponding elements in $N$. More formally, we have the following lemma, which follows easily by choosing $A$ to be a generating set of $G$.

\begin{lemma}\label{lem:allrelationalmorphismsusealphabet}
	For any relational morphism, there exists $A$ and $\mu:A^* \to M$, $\nu:A^* \to N$ such that $\mu$ is surjective and $\tau = \nu \circ \mu^{-1}$, as in the following diagram:
\begin{equation*}
	\begin{tikzcd}
		& A^*	\arrow[dl,"\mu"] \arrow[dr,"\nu"] 	&
		\\	M \arrow[rr, "\tau"] & & N.
	\end{tikzcd}
\end{equation*}
\end{lemma}
Since any such diagram also gives rise to a relational morphism, this means that the relational morphisms between $M$ and $N$ are exactly given by diagrams of this form.

\subsection{Logic}

We will consider $\FO[<]$, first order logic using the following syntax: 
\begin{equation*}
	\varphi ::= \top \mid \bot \mid a(x) \mid x = y \mid x < y \mid \neg\varphi \mid \varphi \wedge \varphi \mid \varphi \vee \varphi \mid \exists x \varphi.
\end{equation*}
Here $a \in A$ for some fixed alphabet $A$, and $\varphi \in \FO[<]$. We interpret formulae in $\FO[<]$ over words as follows. If $i,j \in \mathbb{N}$, then $u,i,j \vDash x < y$ if and only if $i < j$, and $u, i \vDash a(x)$ if and only if $u[i] = a$. The logical connectives and existential quantifier are interpreted as usual. We use the macro $x \leq y$ to mean $x < y \vee x = y$ and the macro $\forall x \varphi$ to mean $\neg \exists x \neg \varphi$. If $\varphi$ is a formula without free variables over the alphabet $A$, we define $L(\varphi) = \left\{ u \in A^* \mid u \vDash \varphi \right\}$. If $\mathscr{F}$ is a collection of formulae, we say that $L \subseteq A^*$ is \emph{definable in $\mathscr{F}$} if there exists $\varphi \in \mathscr{F}$ such that $L = L(\varphi)$.

In particular, we are interested in $\FO^2[<]$, i.e.\ the fragment of $\FO[<]$ where we only allow the use (and reuse) of two variable names. Thus 
\begin{equation*}
	\exists x : a(x) \wedge \left( \exists y : y > x \wedge b(y) \wedge \left(\exists x : x > y \wedge  c(x) \right)\right)
\end{equation*}
is  allowed in $\FO^2[<]$ whereas
\begin{equation*}
	\exists x : a(x) \wedge \left( \exists y : y > x \wedge b(y) \wedge \left(\exists z : z > x \wedge y > z \wedge c(z) \right)\right)
\end{equation*}
is not. It is well known that $\FO^2[<]$ is a proper fragment of $\FO[<]$. We will primarily be interested in some fragments of $\FO^2[<]$. Consider the syntax
\begin{align*}
	\varphi_0 &::= \top \mid \bot \mid a(x) \mid x=y \mid x<y \mid \neg \varphi_{0} \mid \varphi_0 \vee \varphi_0 \mid \varphi_0 \wedge \varphi_0\\
	\varphi_m &::= \varphi_{m-1} \mid \neg \varphi_{m-1} \mid \varphi_m \vee \varphi_m \mid \varphi_m \wedge \varphi_m \mid \exists x \varphi_m
\end{align*}
The collection of formulae $\varphi_m[<]$ is denoted by $\Sigma^2_m[<]$, the collection of negations of formulae in $\Sigma^2_m[<]$ is $\Pi^2_m[<]$ and the Boolean closure of $\Sigma^2_m[<]$ is $\FO^2_m[<]$. In what follows, we drop the reference to the predicate symbol $<$, and assume it to be understood from context.

\subsection{Ramsey Numbers}

A \emph{graph} is a pair $\mathcal{G} = (V,E)$ where $V$ is a set of \emph{vertices} and $E \subseteq \left\{ S \subseteq 2^V \mid |S| = 2\right\}$ is a set of \emph{edges}. An \emph{edge-coloring} is a map $c: E \to C$ where $C$ is some set of colors. A graph is \emph{complete} if $E = \left\{ S \subseteq 2^V \mid |S| = 2 \right\}$, i.e.\ if there is an edge between any two elements. A set $F \subseteq E$ of edges is \emph{monochrome} if $c(e) = c(e')$ for all $e \in F$. A \emph{triangle} is a set of three distinct edges $e_1, e_2, e_3 \in E$ where $e_i \cap e_j \neq \emptyset$ for $1 \leq i,j \leq 3$. The following theorem is a special case of Ramsey's Theorem~\cite{Ramsey1929plms}.

\begin{theorem}\label{thm:ramseys}
	Let $C$ be a finite set of colours. Then there exists a number $R$, called the \emph{Ramsey number} of $C$ such that any complete graph $\mathcal{G} = (V,E)$ with $R \leq |V|$ contains a monochrome triangle.
\end{theorem}

\subsection{Hierarchies Inside $\DA$}
\label{sec:hierinsideda}

A variety of special importance for this article is $\DA$. This monoid variety has a natural correspondence to $\FO^2$ since a language is definable in the latter if and only if its syntactic monoid is in $\DA$.

We are interested in hierarchies of subvarieties of $\DA$. One important such hierarchy is the \emph{Trotter-Weil hierarchy}. Its original motivation comes from an intimate relation with the hierarchy of bands, but here we give a more explicit definition.

\begin{definition}
	Let $M$ be a monoid, and let $s,t \in M$. Then
	\begin{itemize}
		\item $s \sim_{\mathbf{K}} t$ if for all idempotents $e \in M$, either $ev, eu \Jl e$ or $ev = eu$,
		\item $s \sim_{\mathbf{D}} t$ if for all idempotents $f \in M$, either $vf, uf \Jl f$ or $vf = uf$.
	\end{itemize}
	The join of these relations is $\sim_{\mathbf{KD}}$.
\end{definition}

It is straight-forward to check that these relations are congruences (see e.g.\ \cite{krt68arbib8}).
Let $\mathbf{R}_1 = \mathbf{L}_1 = \mathbf{J}_1$, and let $M \in \mathbf{R}_m$ if $\faktor{M}{\sim_{\mathbf{K}}} \,\in \mathbf{L}_{m-1}$ and $M \in \mathbf{L}_m$ if $\faktor{M}{\sim_{\mathbf{D}}}\, \in \mathbf{R}_{m-1}$. When defining $\mathbf{R}_m$ and $\mathbf{L}_m$ for $m \geq 2$, starting with $\mathbf{J}_1$ yields the same result as starting with $\mathbf{J}$. For our purposes, starting with $\mathbf{J}_1$ is more natural.

The varieties $\mathbf{R}_m$ and $\mathbf{L}_m$ are all contained in $\DA$. Together with their joins and intersections they make up the Trotter-Weil hierarchy shown in Figure~\ref{fig:hierarchies} on page~\pageref{fig:hierarchies}.

There is an intimate connection between the quantifier alternation hierarchy, also shown in Figure \ref{fig:hierarchies}, and the Trotter-Weil hierarchy. Indeed, it was shown by Weil and the second author that the languages definable in $\FO^2_m$ are exactly those whose syntactic monoid is in $\mathbf{R}_{m+1} \cap \mathbf{L}_{m+1}$ \cite{kufleitnerweil2012csl}. Furthermore, combining the results in \cite{KufleitnerLauser2012lncs} and \cite{fleischerkufleitnerlauser2014csr} gives the following proposition.

\begin{proposition}\label{prp:rvlandsigmacappi}
	A language is definable in both $\Sigma^2_m$ and $\Pi^2_m$ if and only if its syntactic monoid is in $\mathbf{R}_m \vee \mathbf{L}_m$.
\end{proposition}

The \emph{corners} of the quantifier alternation hierarchy, $\Sigma^2_{m}$ and $\Pi^2_{m}$, also have algebraic characterizations, given by Fleischer, Kuf\-leitner and Lauser \cite{fleischerkufleitnerlauser2014csr}. We will define these recognizing varieties using the stable relation $\preceq_{\mathbf{KD}}$ introduced by the authors \cite{HenrikssonKufleitner20Arxiv}.

\begin{definition}\label{def:KD}
	Let $M$ be a monoid, and let $s,t \in M$. We say that $s \preceq_{\mathbf{KD}} t$ if for all $x,y \in M$, the following holds:
	\begin{enumerate}[(i)]
	      \item\label{aaa:KD} If $x \Req xty$, then $x \Req xsy$,
	      \item\label{bbb:KD} If $xty \Leq y$, then $xsy \Leq y$,
	      \item\label{ccc:KD} If $x \Req xt$ and $ty \Leq y$, then $xsy \leq xty$.
	\end{enumerate}
	If $u \leq_{\mathbf{KD}} v$ and $v \leq_{\mathbf{KD}} u$, we say that $u \equiv_{\mathbf{KD}} v$.\footnote{The name $\preceq_{\mathbf{KD}}$ was originally inspired by the relation $\sim_{\mathbf{KD}}$ since they share some properties. However, it should be noted that the relation $\equiv_{\mathbf{KD}}$ is not the same as $\sim_{\mathbf{KD}}$. As a counter example, note the syntactic monoid of $a^+b^+cA^*da^+b^+$ where the equivalence class of $ab$ is $\equiv_{\mathbf{KD}}$-related to all elements in the minimal \greenJ-class, whereas it is not $\sim_{\mathbf{K}}$- or $\sim_{\mathbf{D}}$-related to anything.}
\end{definition}

Let $\mathbf{Si}_1 = \mathbf{J}^+ = \intp{1 \leq z}$ and let $M  \in \mathbf{Si}_m$ if $\faktor{M}{\preceq_{\mathbf{KD}}} \, \in \mathbf{Si}_{m-1}$. For every $m$, the collection $\mathbf{Si}_m$ is a positive variety. A language is definable in $\Sigma^2_m$ if and only if its syntactic monoid is in $\mathbf{Si}_m$. We say that an ordered monoid $M$ is in $\Pivar_m$ if and only if $M$ with the order reversed, is in $\mathbf{Si}_m$. It is clear that a language is definable in $\Pi^2_m$ if and only if its syntactic monoid is in $\Pivar_m$.

The following property of $\DA$ is standard (see e.g.~\cite{diekertgastinkufleitner2008ijfcs}), and is used throughout the article.

\begin{lemma}\label{lem:daproperty}
	Let $M \in \DA$ and suppose that $M$ has a content morphism $\alpha$. Let $s,t \in M$ and suppose that $\alpha(s) \leq \alpha(t)$, then $s^{\omega} t s^{\omega} = s^{\omega}$.
\end{lemma}

\section{Conelikes and the Covering Problem}
\label{sec:Conelikes}

In this section, we introduce the main problems of the article, the \emph{separation problem} and the \emph{covering problem}.
Given a variety $\mathcal{V}$, the (asymmetric) separation problem is defined as follows:
\begin{quote}
	Given $L,L' \subseteq A^*$, determine if there is a language $K \in \mathcal{V}$ such that $L \subseteq K$ and $L' \cap K = \emptyset$.
\end{quote}
If there exists such a $K$, we say that $L$ is $\mathcal{V}$-separable from $L'$. The symmetric separation problem is to determine whether both $L$ is $\mathcal{V}$-separable from $L'$ and $L'$ is $\mathcal{V}$-separable from $L$. If $\mathcal{V}$ is a full variety, i.e.\ closed under complements, these two problems are equivalent (just choose $A^*\setminus K$ to separate $L'$ from $L$).

There is a strong connection between the (symmetric) separation and the problem of deciding pointlikes \cite{Almeida1999pms}. In this section, we introduce the more general covering problem, together with a generalization of pointlikes which works well with the asymmetric setting. This generalization, which we call \emph{conelikes}, is folklore. However, to the knowledge of the authors they have not been made precise in the algebraic setting.\footnote{The imprints used by Place and Zeitoun in \cite{PlaceZeitoun2018lmcs} yield is a corresponding object in the language setting}

	Let $\mathbf{K}$ be a set of languages, and $\mathbf{L}$ a finite set of languages.  Then $\mathbf{K}$ is \emph{separating} for $\mathbf{L}$ if for all $K \in \mathbf{K}$, there exists $L' \in \mathbf{L}$ such that $K \cap L' = \emptyset$.
	We only consider situations when $\mathbf{K}$ is a \emph{cover} of some language $L$, i.e.\ such that $L \subseteq \bigcup \mathbf{K}$.

\begin{definition}
	Let $\mathcal{V}$ be a (positive) variety. The \emph{covering problem} for $\mathcal{V}$ is defined as follows: 
	\begin{quote}
	Given $L \subseteq A^*$ and $\mathbf{L} \subseteq 2^{A^*}$ where $\mathbf{L}$ is finite, determine if there is $\mathbf{K} \subseteq \mathcal{V}$ which covers $L$ and is separating for $\mathbf{L}$.
	\end{quote}
\end{definition}

If such a $\mathbf{K}$ exists, we say that $(L,\mathbf{L})$ is $\mathcal{V}$-coverable. For $\mathbf{L} = \left\{ L'\right\}$, this reduces to the separation problem. Whenever $\mathcal{V}$ is a full variety, it is equivalent to answer the covering problem for $(L,\mathbf{L})$ and $(A^*,\left\{ L \right\} \cup \mathbf{L})$ \cite{PlaceZeitoun2018lmcs}, and for regular languages this is in turn equivalent to computing the $\mathbf{V}$-pointlikes of a finite monoid recognizing all languages of $\mathbf{L}$ \cite{Almeida1999pms}.

\begin{example}
Consider the variety $\mathcal{J}^+$ of languages whose syntactic monoid is in $\mathbf{J}^+$, and let $L = (ab)^+$, $L_1 = b(ab)^*$, $L_2 = (ab)^*a$. We note that in order to cover $L$, we need the language $A^*aA^*bA^*$. However, this language also contains words from $L_1$ and $L_2$, showing that $(L,\left\{ L_1,L_2 \right\})$ is not $\mathcal{J}^+$-coverable.
\end{example}

\begin{example}
	Suppose that $L \in \mathcal{V}$ can be written as a disjoint union of a finite number of sets $\mathbf{K} \subseteq \mathcal{V}$. Suppose that $\mathbf{K}$ has at least two elements. Then $\mathbf{K}$ is separating for itself (this is the case for all sets with magnitude greater than one), and thus $(L,\mathbf{K})$ is $\mathcal{V}$-coverable. However, it is clear that $L$ is not $\mathcal{V}$-separable from any $L_i \in \mathbf{K}$. 
\end{example}

We want to use algebraic methods to solve the covering problem. However, the covering problem is relevant for positive varieties, whereas pointlikes do not take orders into account.
This motivates the following generalization of pointlikes.

\begin{definition}\label{def:Conelikes}
	Let $M$ be a monoid, and let $\tau: M \to N$ be a relational morphism. For $s \in M$, $S \subseteq M$, we say that $(s,S)$ is \emph{conelike} with respect to $\tau$ if there exists an element $x \in \tau(s)$ such that $S \subseteq \tau^{-1}(\uparrow x)$. We call $x$ a \emph{witness} of $(s,S)$ being conelike. As with pointlikes we say that a pair $(s,S)$ is conelike with respect to a variety $\mathbf{V}$ if it is conelike for any $\tau: M \to N \in \mathbf{V}$. We denote by $\cone{\mathbf{\tau}}{M}$ the conelikes of $M$ with respect to $\tau$, and by $\cone{\mathbf{V}}{M}$ the conelikes of $M$ with respect to $\mathbf{V}$.
\end{definition}

Note that if $N$ is unordered, we can define an order $u \leq v$ if and only if $u = v$. In this case, a pair $(s,S)$ is conelike if and only if $S$ is pointlike and $s \in S$.
In particular, this means that for non-positive varieties, calculating the pointlikes and the conelikes is the same problem.

The concept of pointlikes is in general not expressive enough to solve the covering problem. However, it is still possible to define pointlikes for a variety of ordered monoids, and such pointlikes are used throughout the article. Note that if $S \subseteq M$ is pointlike with respect to some variety of ordered monoids, then $(s,S)$ is conelike for any $s \in S$.

For regular languages, which is what is considered in this contribution, the problem of finding pointlikes is equivalent to the covering problem. We require the following two lemmas. The first is standard when dealing with pointlikes (see e.g.~\cite{Almeida1999pms}) and shows that when finding the optimal conelikes, we need only consider relational morphisms through a fixed alphabet $A$.

\begin{lemma}\label{lem:FixedAlphabet}
	Let $M$ be a monoid, and $\mu: A^* \to M$ a surjective morphism. Let $\mathbf{V}$ be a variety. Then $(s,S) \in M \times 2^M$ is conelike with respect to $\mathbf{V}$ if and only if $(s,S)$ is conelike with respect to all $\relmorsmall{\tau}{A^*}M{N \in \mathbf{V}}\mu\nu$ where $\nu$ is a homomorphism.
\end{lemma}

\begin{proof}
	It is clear that in order to be conelike with respect to $\mathbf{V}$, it must in particular be conelike with respect to the morphisms factoring through $A^*$. For the other direction, suppose $(s,S)$ is conelike with respect to all relational morphisms on the desired form, and let $\rho: M \to N \in \mathbf{V}$ be arbitrary. By Lemma \ref{lem:allrelationalmorphismsusealphabet} we can factor this as $\relmorsmall{\rho}{B^*}MN{f}{g}$ where $f$ and $g$ are homomorphisms, and $\mu'$ is surjective.
	
	We define the homomorphism $h: A^* \to B^*$ as follows. For each $a \in A$, choose $u \in B^*$ such that $f(u) = \mu(a)$ and extend this to a homomorphism. The composition $\varphi^{-1}: g \circ h \circ h^{-1} \circ f^{-1}$ is a relational morphism and factors through $A^*$ in the desired way. Thus, there exists $x \in N$ such that $x \in \varphi(s)$ and $S \subseteq \varphi^{-1}(\uparrow x)$. This in particular implies that $x \in \rho(s)$ and $S \subseteq \rho^{-1}(\uparrow x)$, showing that $(s,S)$ is conelike with respect to $\rho$. Since $\rho$ was arbitrary with co-domain in $\mathbf{V}$, the result follows.
\end{proof}

The next lemma shows that for all varieties $\mathbf{V}$ and monoids $M$, there is a relational morphism $\tau : M \to N \in \mathbf{V}$ which is \emph{optimal} for determining conelikes. Furthermore, this relational morphism can also be assumed to factor through $A^*$.

\begin{lemma}\label{lem:representativetau}
	Let $\mathbf{V}$ be a variety, and let $M$ be a finite monoid. Then there exists $\tau: M{N \in \mathbf{V}}\mu\nu$ such that $\cone{\mathbf{V}}{M} = \cone{\tau}{M}$. Furthermore, if $\mu: A^* \to M$ is a surjective homomorphism, we can assume that $\tau$ can be written as $\relmorsmall\tau{A^*}MN\mu\nu$ where $\mu$ and $\nu$ are homomorphisms and $\mu$ is surjective.
\end{lemma}

\begin{proof}
	Let $\tau_1: M \to N_1$ be arbitrary. We will define a chain $\tau_1, \dots, \tau_n$ where $\cone{\tau_{i+1}}{M} \subsetneq \cone{\tau_{i}}{M}$, using the following procedure. If $\cone{\tau_i}{M} = \cone{\mathbf{V}}{M}$, we set $i = n$ and are done. If not, there exists $\tau': M \to N' \in \mathbf{V}$ such that $\cone{\tau_i}{M} \not \subseteq \cone{\tau'}{M}$. We define $\tau_{i+1}: M \to N_i \times N'$ by $\tau_{i+1}(s) = \{(t,t')\}_{t \in \tau_i(s), t' \in \tau'(s)}$. It is clear that this is a relational morphism, and that $(s,S) \in \cone{\tau_{i+1}}{M}$ if and only if $(s,S) \in \cone{\tau_{i}}{M} \cap \cone{\tau'}{M}$. By choice of $\tau'$, this shows that $\tau_{i+1}$ has the desired properties. Since $M$ is finite, this process must eventually stop giving the desired result.

	For the final part of the lemma, suppose that $\tau_n$ has the desired properties. Using the same argument as in Lemma \ref{lem:FixedAlphabet}, we get a relational morphism $\relmorsmall\tau{A^*}MN\mu\nu$ such that every conelike with respect to $\tau$ is also conelike with respect to $\tau_n$ and thus also with respect to $\mathbf{V}$.
\end{proof}

\begin{proposition}\label{prp:conelikesandpointedcoverrelation}
	Let $\mathbf{L} = \left\{ L_i \right\}$ be a finite collection of regular languages, and for each $L_i$ let $\mu_i: A^{*} \to M_i$ be its syntactic monoid. Let $\mu: A^* \to M_1 \times \dots \times M_n$ be defined by $\mu(a) = (\mu_1(a),\dots,\mu_n(a))$ and let $M = \mu(A^*)$. Let $\mathbf{V}$ be a variety of monoids recognizing a variety $\mathcal{V}$ of languages. Then the following are equivalent
	\begin{enumerate}[(i)]
		\item $(L,\mathbf{L})$ is \textbf{not} $\mathcal{V}$-coverable,
		\item there exists a conelike $(s,S)$ with respect to $\mathbf{V}$ such that $L \cap \mu^{-1}(s) \neq \emptyset$ and for all $L' \in S$ there exists $s' \in S$ such that $L' \cap \mu^{-1}(s') \neq \emptyset$.\label{bbb:conelikesandpointedcoverrelation}
	\end{enumerate}
\end{proposition}

\begin{proof}
	Let $\relmorsmall\tau{A^*}MN\mu\nu$ be a relational morphism with $N \in \mathbf{V}$ chosen to have the property in Lemma \ref{lem:representativetau}. Let
	\begin{equation*}
		\mathbf{P} = \left\{ (L,\nu^{-1}(\uparrow x)) \mid x \in \nu(L) \right\}.
	\end{equation*}
	It is clear that the set $\bigcup_{x \in \nu(L)}\nu^{-1}(\uparrow x)$ is a cover of $L$. If $(L,\mathbf{L})$ is not $\mathcal{V}$-coverable, then there exists $x \in \nu(L)$ such that for all $L' \in \mathbf{L}$, we have $L' \cap \nu^{-1}(x) \neq \emptyset$. Choose $s \in \tau^{-1}(x) \cap \mu(L)$ arbitrary; such $s$ exist since we only consider $x$ in the image of $L$. Then $(s,\tau^{-1}(\uparrow x))$ is a conelike with property \itref{bbb:conelikesandpointedcoverrelation}.

	For the other direction, let $\mathbf{K} \subseteq \mathcal{V}$ cover $L$ and be separating for $\mathbf{L}$ and let $\nu: A^* \to N$ recognize all $K_i \in \mathbf{K}$. Let $\relmorsmall\tau{A^*}MN\mu\nu$ be the natural relational morphism.
	For contradiction, assume that there is a conelike with property \itref{bbb:conelikesandpointedcoverrelation}. Then we find $u \in L$ and $v_i \in L_i$ for each $L_i \in U$, such that $(\mu(u),\left\{ \mu(v_i) \right\}_i)$ is conelike with respect to $\tau$. In particular, we can assume the $u$ and $v_i$ to be the representatives for which $\nu(v_i) \in \uparrow \nu(u)$. Now, $\uparrow \nu(u)$ intersects every language in $\mathbf{L}$, and thus $\mathbf{K}$ is not separating for $\mathbf{L}$.
\end{proof}

\begin{example}\label{exm:PointlikeExample}
	Let us return to our example with the languages $(ab)^+$, $b(ab)^*$ and $(ab)^*a$. These are all recognized by the syntactic monoid of $(ab)^+$. The above proposition tells us that $\left( ab,\left\{ b,a \right\} \right)$ is conelike, where we identify the languages with representatives of their equivalence classes. In particular, an optimal relational morphism is the natural morphism into the syntactic monoid of $A^*aA^*bA^*$.
\end{example}

Before leaving the topic of conelikes, we introduce a common tool for determining pointlikes and solving the covering problem (see e.g.~\cite{PlaceZeitoun2018lmcs}). The idea is to construct sets of subsets of $M$ which have closure properties analogous to those of pointlikes and conelikes.

\begin{definition}
	Given a monoid $M$, a subset of $2^M$ is \emph{closed} if it contains the singletons and is closed under multiplication and subsets. Similarly, a set $\mathcal{C} \subseteq M \times 2^{M}$ is closed if it has the following closure properties:
	\begin{itemize}
		\item $(s,\left\{ s \right\}) \in \mathcal{C}$ for all elements $s \in M$,
		\item $(s,S), (t,T) \in \mathcal{C}$ implies $(st,ST) \in \mathcal{C}$,
		\item $(s,S) \in \mathcal{C}$ implies $(s,S') \in \mathcal{C}$ for all subsets $S' \subseteq S$.
	\end{itemize}
\end{definition}

In Sections \ref{sec:decidingconelikes} to \ref{sec:DifficultDirection}, we introduce computable closed sets for all varieties of interest and show that these sets coincide with the pointlikes and conelikes.

\section{A Framework for Ranker Comparisons}
\label{sec:rankercomp}

One of the main techniques in this paper is ranker comparisons. This concept has close connections to fragments of $\FO^2$, a connection we explore in Section \ref{sec:RankerCompHierarchy} (see.~\cite{WeisImmerman2009lmcs,Lauser2014thesis}). In this section, we introduce a general framework and give sufficient conditions for instances of this framework to define a monoid.

\begin{definition}
	Let $A$ be some alphabet. A \emph{ranker} over $A$ is a nonempty word over $\left\{ \mathsf{X}_a, \mathsf{Y}_a \right\}_{a \in A}$, which can be interpreted as a partial function from $A^*$ to $\mathbb{N}$. The interpretation is defined inductively as follows:
	\begin{itemize}
		\item $\mathsf{X}_a(u) = \inf\left\{ n \in \mathbb{N} \mid u[n] = a \right\}$ if the infimum is finite, and undefined otherwise,
		\item $\mathsf{Y}_a(u) = \sup\left\{ n \in \mathbb{N} \mid u[n] = a \right\}$ if the supremum is finite, and undefined otherwise,
		\item $r\mathsf{X}_a(u) = \inf\left\{ n \in \mathbb{N} \mid n > r(u), u[n] = a \right\}$ if $r(u)$ is defined and the infimum is finite, and undefined otherwise,
		\item $r\mathsf{Y}_a(u) = \sup\left\{ n \in \mathbb{N} \mid n < r(u), u[n] = a \right\}$ if $r(u)$ is defined and the supremum is finite, and undefined otherwise.
	\end{itemize}
\end{definition}

Note that we read rankers from left to right (as opposed to function composition). Thus $\mathsf{X}_a\mathsf{Y}_b(bab) = 1$, whereas $\mathsf{Y}_b\mathsf{X}_a(bab)$ is undefined. If $p = a_1 \cdots a_n$, we define $\mathsf{X}_p = \mathsf{X}_{a_1} \cdots \mathsf{X}_{a_n}$ and $\mathsf{Y}_p = \mathsf{Y}_{a_1} \cdots \mathsf{Y}_{a_n}$ for compactness.

We define the following collections of rankers:
	\begin{align*}
	R^X_{1} &= \left\{ X_a \right\}_{a \in A}^+,
	&R^Y_{1} &= \left\{ Y_a \right\}_{a \in A}^+,
	&R^X_{m+1} &= \left\{ X_a \right\}_{a \in A}^* R^Y_{m},
	&R^Y_{m+1} &= \left\{ Y_a \right\}_{a \in A}^* R^X_{m},
	\end{align*}
	Here the juxtaposition on the left denotes concatenation. We furthermore define $R_{m} = R^X_{m} \cup R^Y_{m}$, and $R = \bigcup_m R_m$. Note that these sets depend on the alphabet $A$ although this dependence is not written out explicitly. We will always assume that the alphabet is clear from context.
Given a ranker $r$, the \emph{alternation depth} of $r$ is the smallest $m$ such that $r \in R_m$. The \emph{depth} of $r$ is the length of $r$ as a word.

Since rankers are $\mathbb{N}$-valued functions, there is a natural way of comparing them given a speficied word $u$. In other words, given $u \in A^*$ and rankers $r,s$, we are interested in whether $r(u) \leq s(u)$ and $r(u) < s(u)$ hold. The following definition introduces a general framework, inspired by the comparisons of Weis and Immerman \cite{WeisImmerman2009lmcs} and Lauser \cite{Lauser2014thesis}.

\begin{definition}\label{def:comparisonssets}
	Let $A$ be some alphabet, and let $\mathscr{C} \subseteq R \times R$ be some set of pairs of rankers over $A$. We define $[\mathscr{C}] = \bigcup_{(r,s) \in \mathscr{C}}\left\{ r,s \right\}$, i.e.\ the set of rankers that occurs on some position in some pair of $\mathscr{C}$. We say that $u \leq^{\mathscr{C}} v$ if:
	\begin{enumerate}[\itshape(i)]
		\item The same set of rankers in $[\mathscr{C}]$ are defined on $u$ and $v$,\label{aaa:comparisonssets}
		\item For each $(r,s) \in \mathscr{R}$ such that $r$ and $s$ are defined on $u$ and $v$, we have $r(u) \leq s(u) \Rightarrow r(v) \leq s(v)$, and $r(u) < s(u) \Rightarrow r(v) < s(v)$.
	\end{enumerate}
	If $u \leq^{\mathscr{C}} v$ and $v \leq^{\mathscr{C}} u$, we say that $u \equiv^{\mathscr{C}} v$. 
\end{definition}

For rankers $r$,$s$, and words $u,v \in A^*$ we have $r(u) \leq s(u) \Rightarrow r(v) \leq s(v)$  if and only if $s(v) < r(v) \Rightarrow s(u) < r(u)$. In particular, this means that if $\mathscr{C}$ is symmetric, i.e.\ $(r,s) \in \mathscr{C} \Leftrightarrow (s,r) \in \mathscr{C}$, then $\leq^{\mathscr{C}}$ and $\equiv^{\mathscr{C}}$ are equivalent relations. 
	Note that for a certain choice of symmetric $\mathscr{C}$ we get the relation introduced in \cite{WeisImmerman2009lmcs}.
	We say that a language $L$ is \emph{definable} by $\mathscr{C}$ if $L$ is an ideal under the relation $\leq^{\mathscr{C}}$. Furthermore, we say that a language is an \emph{$\mathscr{C}$-set} if it is a subset of such an ideal. 

For a language $A$ and some sets of rankers $\mathscr{C} \subseteq R \times R$, we want to consider the monoid $\faktor{A^*}{\leq^{\mathscr{C}}}$, which is a well defined monoid only when $\leq^{\mathscr{C}}$ is stable. For general $\mathscr{C}$, this is not the case. However, Proposition \ref{prp:PreorderIsStable} provides a large class of sets for which it does hold.\footnote{As an example on when it does not hold, consider the singleton $\left\{ (\mathsf{X}_{aa},\mathsf{Y}_{aa}) \right\}$. We note that neither $\mathsf{X}_{aa}$ nor $\mathsf{Y}_{aa}$ are defined on $\varepsilon$ nor on $a$. Thus $\varepsilon \leq^{\mathscr{C}} a$. However, $a \not\leq^{\mathscr{C}} aa$. The following proposition gives a condition on $\mathscr{C}$ which implies that $\leq^{\mathscr{C}}$ is stable.}

\begin{proposition}\label{prp:PreorderIsStable}
	Let $R$ be some collection of rankers and let $\mathscr{C} \subseteq R \times R$ be closed under subwords, i.e.\ be such that $(r,s) \in \mathscr{C}$ implies $(r',s') \in \mathscr{C}$ for any subwords $r'$ of $r$ and $s'$ of $s$. Then the preorder $\leq^{\mathscr{C}}$ is stable.
\end{proposition}

\begin{proof}
	By symmetry, it is enough to show that $u \leq^{\mathscr{C}} v$ implies $xu \leq^{\mathscr{C}} xv$.
	Suppose $r \in [\mathscr{C}]$ is defined on $xu$. Without loss of generality, we assume that $r$ starts with an $\mathsf{X}$-modality.
	
	We will factor $r = s_1t_1 \dots s_nt_n$ in the following way; let $s_1$ be the longest (possibly empty) prefix of $r$ which is defined on $x$, and factor $r = s_1r'$. Next, let $t_1$ be the longest prefix of $r'$ which is defined on $u$, and continue this process. Since $t_i$ is defined on $u$, it is also defined on $v$, and since no longer factor is defined on $u$, no longer factor is defined on $v$. Thus, the factorization of $r$ would be the same if taken with respect to $xv$, and in particular, $r$ is defined on $xv$.

	Suppose next that $r(xu) < r'(xu)$. We factor $r = s_1t_1 \dots s_kt_k$ and $r' = s_1't_1' \dots s_{\ell}'t_{\ell}'$ as before where we allow $s_1$,$s'_1$, $t_k$ and $t_{\ell}'$ to be empty. By the above argument, the $t_i$ and $t_j'$ are also the longest factors being defined also on $v$.
	Suppose that $t_k$ is empty, but $t'_{\ell}$ is not. Then $r(xv) \leq |x| < |x| + 1 \leq r'(xv)$. Note that we can not have $t_k$ nonempty while $t_{\ell}'$ is empty since this would contradict $r(xu) < r'(xu)$. Finally, if $t_k$ and $t_{\ell}'$ are both nonempty. Then $t_k(u) < t_{\ell}'(u)$, and we get $r(xv) = |x| + t_k(v) < |x| + t'_{\ell}(v) = r'(xv)$. The case when both $t_k$ and $t_{\ell}'$ are empty is handled similarly. Thus, we get $r(xv) < r'(xv)$. The case dealing with the non-strict order is analogous.
\end{proof}

If $\mathscr{C}$ furthermore is finite, then $\faktor{A^*}{\leq^{\mathscr{C}}}$ is a finite monoid. This monoid can be constructed explicitly; given representatives $u$ and $v$ for some elements, one can check which ranker comparisons $uv$ satisfy.

We also mention the two following lemmas which are trivial and have been written down without a proof.

\begin{lemma}\label{lem:MoveLastAlternation}
	Let $r = r'\mathsf{X}_{p}$ and $s$ be rankers. If $r(u)$ and $s(u)$ are defined, then
	\begin{align*}
		r(u) \leq s(u) &\Leftrightarrow r'(u) \leq s\mathsf{Y}_{\overline{p}}(u)
		& r(u) < s(u) &\Leftrightarrow r'(u) < s\mathsf{Y}_{\overline{p}}(u)
		\\ s(u) \leq r(u) &\Leftrightarrow s\mathsf{Y}_{\overline{p}}(u) \leq r'(u)
		& s(u) < r(u) &\Leftrightarrow s\mathsf{Y}_{\overline{p}}(u) < r'(u).
	\end{align*}
	where in the second line the implications only holds if $s\mathsf{Y}_{\overline{p}}(u)$ is defined. In the first line, $s\mathsf{Y}_{\overline{p}}(u)$ being defined is also implied. Symmetrically, if $r(u) = r'\mathsf{Y}_{u}(u)$, and $r(u)$ and $s(u)$ are defined, then
	\begin{align*}
		r(u) \leq s(u) &\Leftrightarrow r'(u) \leq s\mathsf{X}_{\overline{p}}(u)
		& r(u) < s(u) &\Leftrightarrow r'(u) < s\mathsf{X}_{\overline{p}}(u)
		\\ s(u) \leq r(u) &\Leftrightarrow s\mathsf{X}_{\overline{p}}(u) \leq r'(u)
		& s(u) < r(u) &\Leftrightarrow s\mathsf{X}_{\overline{p}}(u) < r'(u).
	\end{align*}
	where this time definedness is only implied in the second line.
\end{lemma}

\begin{lemma}\label{lem:DefinedOnFactor}
	Let $u = xavby$ where $u,v,x,y \in A^*$, $a,b \in A$. If there are rankers $r,s$ such that $r(u) = |xa|$, $s(u) = |xavb|$, then a ranker $t$ starting with an $\mathsf{X}$-modality is defined on $v$ if and only if $rt$ is defined on $u$ and $r(u) < rt'(u) < s(u)$ for all prefixes $t'$ of $t$. 
	Symmetrically, if $t$ starts with a $\mathsf{Y}$-modality, then $t$ is defined on $v$ if and only if $st$ is defined on $u$ and $r(u) < st'(u) < s(u)$ for all prefixes $t'$ of $t$. 
\end{lemma}

\section{The Ranker Comparison Hierarchy}
\label{sec:RankerCompHierarchy}

Rankers and ranker comparisons have a long tradition in the study of fragments of $\FO^2$. Indeed, rankers were first introduced by Schwentick, Th\'erien and Vollmer as a characterization of $\FO^2$ itself \cite{SchwentickTherienVollmer2001dlt}. Ranker comparisons were used by Weis and Immerman to give a characterization of the languages definable in $\FO^2_m$ \cite{WeisImmerman2009lmcs}; this was later expanded to the full alternation hierarchy in the PhD thesis of Lauser \cite{Lauser2014thesis}.
A ranker characterization of the corners of the Trotter-Weil hierarchy is also known, using so called condensed rankers \cite{KufleitnerWeil2012lmcs}.

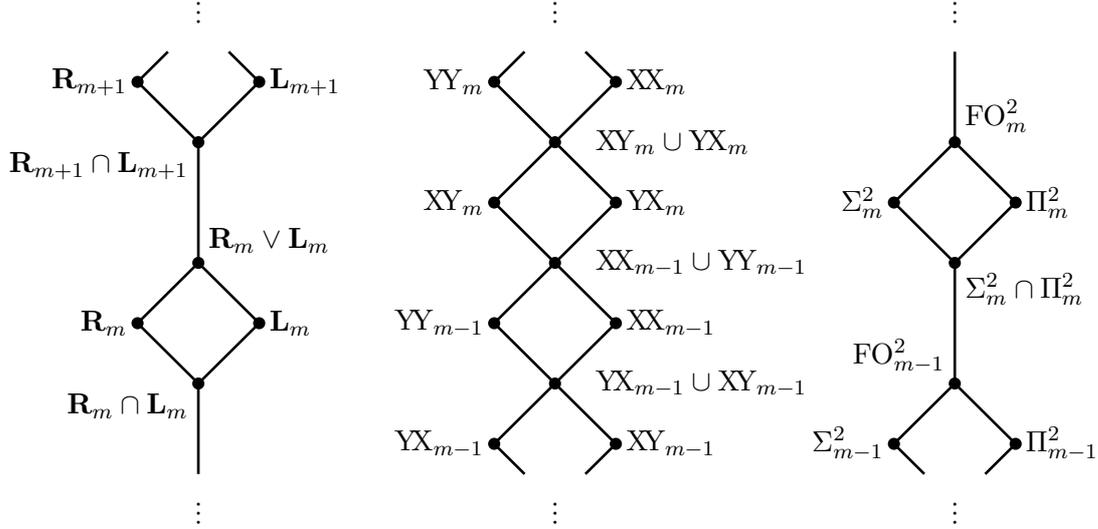
\begin{figure}
\centering
\begin{subfigure}{0.3 \textwidth}
\begin{tikzpicture}[scale = 0.4]
	\fill (0,0) circle [radius=0.05];
	\fill (0,-0.3) circle [radius=0.05];
	\fill (0,-0.6) circle [radius=0.05];
	\draw [line width = 1] (0,1) -- (0,4);
	\fill (0,4) circle [radius=0.2] node [below left] {$\mathbf{R}_m \cap \mathbf{L}_m$};	
	\fill (2,6) circle [radius=0.2] node [right] {$\mathbf{L}_m$};	
	\fill (-2,6) circle [radius=0.2] node [left] {$\mathbf{R}_m$};	
	\draw [line width = 1] (0,4) -- (2,6);
	\draw [line width = 1] (0,4) -- (-2,6);
	\draw [line width = 1] (0,8) -- (2,6);
	\draw [line width = 1] (0,8) -- (-2,6);
	\fill (0,8) circle [radius=0.2] node [above right] {$\mathbf{R}_m \vee \mathbf{L}_m$};
	\draw [line width = 1] (0,8) -- (0,12);
	\fill (0,12) circle [radius=0.2] node [below left] {$\mathbf{R}_{m+1} \cap \mathbf{L}_{m+1}$};	
	\fill (2,14) circle [radius=0.2] node [right] {$\mathbf{L}_{m+1}$};	
	\fill (-2,14) circle [radius=0.2] node [left] {$\mathbf{R}_{m+1}$};	
	\draw [line width = 1] (0,12) -- (2,14);
	\draw [line width = 1] (0,12) -- (-2,14);
	\draw [line width = 1] (1,15) -- (2,14);
	\draw [line width = 1] (-1,15) -- (-2,14);
	\fill (0,16) circle [radius=0.05];
	\fill (0,16.3) circle [radius=0.05];
	\fill (0,16.6) circle [radius=0.05];
\end{tikzpicture}	
\end{subfigure}\;\;
\begin{subfigure}{0.3\textwidth}
\begin{tikzpicture}[scale = 0.4]
	\fill (0,0) circle [radius=0.05];
	\fill (0,-0.3) circle [radius=0.05];
	\fill (0,-0.6) circle [radius=0.05];
	\fill (2,2) circle [radius=0.2] node [right] {$\XY_{m-1}$};	
	\fill (-2,2) circle [radius=0.2] node [left] {$\YX_{m-1}$};	
	\draw [line width = 1] (1,1) -- (2,2);
	\draw [line width = 1] (-1,1) -- (-2,2);
	\draw [line width = 1] (0,4) -- (2,2);
	\draw [line width = 1] (0,4) -- (-2,2);
	\fill (0,4) circle [radius=0.2] node [right = 0.4] {$\YX_{m-1} \cup \XY_{m-1}$};	
	\fill (2,6) circle [radius=0.2] node [right] {$\XX_{m-1}$};	
	\fill (-2,6) circle [radius=0.2] node [left] {$\YY_{m-1}$};	
	\draw [line width = 1] (0,4) -- (2,6);
	\draw [line width = 1] (0,4) -- (-2,6);
	\draw [line width = 1] (0,8) -- (2,6);
	\draw [line width = 1] (0,8) -- (-2,6);
	\fill (0,8) circle [radius=0.2] node [right = 0.4] {$\XX_{m-1} \cup \YY_{m-1}$};	
	\fill (2,10) circle [radius=0.2] node [right] {$\YX_{m}$};	
	\fill (-2,10) circle [radius=0.2] node [left] {$\XY_{m}$};	
	\draw [line width = 1] (0,8) -- (2,10);
	\draw [line width = 1] (0,8) -- (-2,10);
	\draw [line width = 1] (0,12) -- (2,10);
	\draw [line width = 1] (0,12) -- (-2,10);
	\fill (0,12) circle [radius=0.2] node [right = 0.4] {$\XY_{m} \cup \YX_{m}$};	
	\fill (2,14) circle [radius=0.2] node [right] {$\XX_{m}$};	
	\fill (-2,14) circle [radius=0.2] node [left] {$\YY_{m}$};	
	\draw [line width = 1] (0,12) -- (2,14);
	\draw [line width = 1] (0,12) -- (-2,14);
	\draw [line width = 1] (1,15) -2 (2,14);
	\draw [line width = 1] (-1,15) -- (-2,14);
	\fill (0,16) circle [radius=0.05];
	\fill (0,16.3) circle [radius=0.05];
	\fill (0,16.6) circle [radius=0.05];
\end{tikzpicture}
\end{subfigure}\;\;\;\;\;\;
\begin{subfigure}{0.3\textwidth}
\begin{tikzpicture}[scale = 0.4]
	\fill (0,0) circle [radius=0.05];
	\fill (0,-0.3) circle [radius=0.05];
	\fill (0,-0.6) circle [radius=0.05];
	\fill (-2,2) circle [radius=0.2] node [left] {$\Sigma^2_{m-1}$};	
	\fill (2,2) circle [radius=0.2] node [right] {$\Pi^2_{m-1}$};	
	\draw [line width = 1] (1,1) -- (2,2);
	\draw [line width = 1] (-1,1) -- (-2,2);
	\draw [line width = 1] (0,4) -- (2,2);
	\draw [line width = 1] (0,4) -- (-2,2);
	\fill (0,4) circle [radius=0.2] node [above left] {$\FO^2_{m-1}$};	
	\draw [line width = 1] (0,4) -- (0,8);
	\fill (0,8) circle [radius=0.2] node [below right] {$\Sigma^2_{m} \cap \Pi^2_{m}$};	
	\fill (-2,10) circle [radius=0.2] node [left] {$\Sigma^2_{m}$};	
	\fill (2,10) circle [radius=0.2] node [right] {$\Pi^2_{m}$};	
	\draw [line width = 1] (0,8) -- (2,10);
	\draw [line width = 1] (0,8) -- (-2,10);
	\draw [line width = 1] (0,12) -- (2,10);
	\draw [line width = 1] (0,12) -- (-2,10);
	\fill (0,12) circle [radius=0.2] node [above right] {$\FO^2_{m}$};	
	\draw [line width = 1] (0,12) -- (0,15);
	\fill (0,16) circle [radius=0.05];
	\fill (0,16.3) circle [radius=0.05];
	\fill (0,16.6) circle [radius=0.05];
\end{tikzpicture}
\end{subfigure}
\
\caption{The Ranker Comparison Hierarchy surrounded by the Trotter-Weil hierarchy (left) and the Quantifier Alternation Hierarchy (right)}\label{fig:hierarchies}
\end{figure}

In this section, we place these results into our general framework. In particular, we rephrase the characterization of the Trotter-Weil corners in terms of ranker comparisons. This leads to a natural hierarchy containing both the Trotter-Weil and quantifier alternation hierarchies: the \emph{ranker comparison hierarchy}, shown in Figure \ref{fig:hierarchies}.

The levels of the ranker comparison hierarchy are built using the following collections of ranker comparisons. We note that they are finite and closed under subwords, and thus define finite monoids by Proposition \ref{prp:PreorderIsStable}. For $m \geq 1$:
\begin{align*}
	 \XX_{m,n} & = \left\{ (r,s) \in  R^X_{m} \times R^X_{m} \mid |r|,|s| \leq n \right\}, &
	 \YY_{m,n} & = \left\{ (r,s) \in  R^Y_{m} \times R^Y_{m} \mid |r|,|s| \leq n \right\}
\end{align*}
and for $m \geq 2$:
\begin{align*}
	\XY_{m,n} & = \left\{ (r,s) \in  R^X_m \times R^Y_m \mid |r|,|s| \leq n \right\},
	& \YX_{m,n} & = \left\{ (r,s) \in  R^Y_m \times R^X_m \mid |r|,|s| \leq n \right\}.
\end{align*}
We also consider unions of these sets. We will use the notation $\leq^{\XY}_{m,n}$ instead of $\leq^{\XY_{m,n}}$ and similarly for the other sets. Note that we have $u \leq^{\XY}_{m,n} v$ if and only if $v \leq^{\YX}_{m,n} u$ and $v \leq^{\XY \cup \YX}_{m,n} u$ if and only if both $v \leq^{\XY}_{m,n} u$ and $v \leq^{\YX}_{m,n} u$.\footnote{Note that although the relation $\leq^{\XY \cup \YX}_{m,n}$ is similar to the relation introduced by Weis and Immerman \cite{WeisImmerman2009lmcs}, there is a slight different regarding the variable $n$. The length of rankers allowed by Weis and Immerman is made to ensure correspondence with the depth of formulae in $\FO^2$. The results of this article does not consider depths of $\FO^2$ formulae, and thus this difference is not important here.} We also give the following names for the induced monoids.

\begin{definition}
	We define
	\begin{align*}
		N^{\XX}_{m,n} & = \faktor{A^*}{\leq^{\XX}_{m,n}}
		&N^{\YY}_{m,n} & = \faktor{A^*}{\leq^{\YY}_{m,n}}
		& N^{\XX \cup \YY}_{m,n} & = \faktor{A^*}{\leq^{\XX \cup \YY}_{m,n}}
		\\ N^{\XY}_{m,n} & = \faktor{A^*}{\leq^{\XY}_{m,n}}
		&N^{\YX}_{m,n} & = \faktor{A^*}{\leq^{\YX}_{m,n}}
		&N^{\XY \cup \YX}_{m,n} & = \faktor{A^*}{\leq^{\XY \cup \YX}_{m,n}}.
	\end{align*}
\end{definition}

The fragment $\FO^2_{1,n}$ is characterized by existence of subwords. Since our framework require the same subwords up to a certain length to be present in both words of interest (condition \itref{aaa:comparisonssets} in Definition \ref{def:comparisonssets}), this means that our framework can not properly handle this case. Therefore, we make the following special definition.

\begin{definition}\label{def:xyspecial}
	Let $A$ be some alphabet, we say that $u \leq^{\XY}_{1,n} v$ if any ranker in $R^X_{1,2n} \cup R^Y_{1,2n}$ which is defined on $u$ is also defined on $v$. Equivalently, $u \leq^{\XY}_{1,n} v$ if every subword of length $2n$ which exists in $u$ also exists in $v$. We say that $u \leq^{\YX}_{1,n} v$ if $v \leq^{\XY}_{1,n} u$ and $u \leq^{\XY \cup \YX}_{1,n} v$ if $u \equiv^{\XY}_{1,n} v$.
\end{definition}

A language is definable by $\XY_{1,n}$ (resp. $\YX_{1,n}$ or $\XY_{1,n} \cup \YX_{1,n}$) if it  is an ideal under the relation $\leq^{\XY}_{1,n}$ (resp. $\leq^{\YX}_{1,n}$ or $\leq^{\YX \cup \YX}_{1,n}$). This does, however, \emph{not} mean that we can interpret $\XY_{1,n}$ as a set of rankers and definability in the sense of Definition \ref{def:comparisonssets}. We also define $N^{\XY}_{1,n}$ and $N^{\YX}_{1,n}$ to be the monoids induced by the respective (stable) preorders.
This special definition ensures that we can define all levels of the quantifier alternation hierarchy using consistent terminology.
Furthermore, it makes the following lemma true for all $m$.

\begin{lemma}\label{lem:subwordsubranker}
	Let $m \geq 2$ and let $u,u' \in A^*$ such that
	\begin{equation*}
		u = xavby \leq^{\YX}_{m,n}  x'av'by' = u'
	\end{equation*}
	where $x,v,y,x',v',y' \in A^*$ and $a,b \in A$. If there exists rankers $\mathsf{X}_{p}$, $\mathsf{Y}_q$ with $|p|,|q| \leq k < n$ such that $\mathsf{X}_p(u) = |xavb|$, $\mathsf{X}_p(u') = |x'av'b|$, $\mathsf{Y}_q(u) = |xa|$ and $\mathsf{Y}_q(u') = |x'a|$, then $v \leq^{\XY}_{m-1,n-k} v'$. Similarly, if $u \leq^{\XY}_{m,n} u'$, then $v \leq^{\YX}_{m-1,n-k} v'$.
\end{lemma}

In other words, the implication of true ranker comparisons carries over from $u$ and $u'$ to $v$ and $v'$ at the cost of one quantifier alternation and the depth needed to reach the positions $a$ and $b$.

\begin{proof}
	We start by considering the case $m \geq 3$, since $m = 2$ needs to be treated specifically. Let $(r,s) \in \XY_{m-1,n-k}$.
	
	We first assume that one of the rankers, say $r$, is defined on $v$ and show that this implies that it is also defined on $v'$. Suppose $r$ starts with an $\mathsf{X}$-modality, and let $r'$ be an arbitrary prefix of $r$. We have $\mathsf{Y}_qr'(u) < \mathsf{X}_p(u)$ which implies $\mathsf{Y}_qr'(u') < \mathsf{X}_p(u')$ since $(\mathsf{Y}_qr',\mathsf{X}_p) \in \YX_{m,n}$.

	We also need to show $\mathsf{Y}_{q}(u') < \mathsf{Y}_qr'(u')$, however, since the alternation depth of $r'$ is possibly $m-1$, the comparison $(\mathsf{Y}_q,\mathsf{Y}_qr')$ might not be in $\YX_{m,n}$. 
	Suppose $r'$ ends with a $\mathsf{Y}$-modality, say $r'  = r''\mathsf{Y}_t$, where again $r''$ has a strictly smaller alternation depth. Then
  \begin{alignat*}{2}
	  \mathsf{Y}_p(u) \leq \mathsf{Y}_qr''\mathsf{Y}_t(u)
	  \quad \Rightarrow \quad & \mathsf{Y}_p\mathsf{X}_{\overline{t}}(u) \leq \mathsf{Y}_qr''(u)
	  &\qquad\quad&\text{By Lemma \ref{lem:MoveLastAlternation}}
    \\ \Rightarrow \quad & \mathsf{Y}_p\mathsf{X}_{\overline{t}}(u') \leq \mathsf{Y}_qr''(u')
    &&\text{since $u \leq^{\YX}_{m,n} u'$}
    \\ \Rightarrow \quad & \mathsf{Y}_p(u') \leq \mathsf{Y}_qr''\mathsf{Y}_t(u')
    &&\text{By Lemma \ref{lem:MoveLastAlternation}}
  \end{alignat*}
  where in the second implication, we rely on the fact that $r''$ has alternation depth at most $m-2$, and thus $Y_qr''$ has alternation depth at most $m-1$.
  If $r'$ ends with an $\mathsf{X}$-modality, we get the same chain of equivalences. Thus we have $\mathsf{Y}_q(u') < \mathsf{Y}_qr'(u') < \mathsf{X}_p(u)$ for all prefixes of $r$, which by Lemma \ref{lem:DefinedOnFactor} implies that $r$ is defined on $v'$. The case when $r$ starts with a $\mathsf{Y}$-modality is symmetric.

	Assume next that $r$ is defined on $v'$, and let $r'$ be an arbitrary prefix. We will use induction on the alternation depth of $r'$ to show that every such prefix is defined on $r$. We consider the case when $r' = r''\mathsf{X}_t$ where $r''$ is either empty or has strictly smaller alternation depth. We also assume $r$ starts with an $\mathsf{X}$-modality. The other cases are similar.
	
	By induction $r''$ is either empty or defined on $v$, and it thus suffices to show $\mathsf{Y}_qr''\mathsf{X}_t(u) < \mathsf{X}_p(u)$. Suppose this is not the case, then
  \begin{alignat*}{2}
	  \mathsf{X}_p(u) \leq \mathsf{Y}_qr''\mathsf{X}_t(u)
	  \quad \Rightarrow \quad & \mathsf{X}_p\mathsf{Y}_{\overline{t}}(u) \leq \mathsf{Y}_qr''(u)
	  &\qquad\quad&\text{By Lemma \ref{lem:MoveLastAlternation}}
    \\ \Rightarrow \quad & \mathsf{X}_p\mathsf{Y}_{\overline{t}}(u') \leq \mathsf{Y}_qr''(u')
    &&\text{since $u \leq^{\YX}_{m,n} u'$}
    \\ \Rightarrow \quad & \mathsf{X}_p(u') \leq \mathsf{Y}_qr''\mathsf{X}_t(u')
    &&\text{By Lemma \ref{lem:MoveLastAlternation}}
  \end{alignat*}
  where in the second implication, we both rely on the fact that $m \geq 3$ and the alternation depth of $r''$ is at most $m-2$. This contradicts $r'$ being defind on $v'$. Thus, we must have $\mathsf{Y}_qr'(u) < \mathsf{X}_p(u)$ which means $r'$ is defined on $v$.
  
  Finally, suppose that $r(u) \leq s(u)$. We suppose $r$ starts with an $\mathsf{X}$-modality and $s$ with a $\mathsf{Y}$-modality. Then $\mathsf{Y}_qr(u) \leq \mathsf{X}_ps(u)$ holds, giving $\mathsf{Y}_qr(u') \leq \mathsf{X}_ps(u')$. Since $r$ and $s$ are defined on $v'$, this implies $r(v') \leq s(v')$. The other cases are similar, using the fact that if $r$ starts with a $\mathsf{Y}$-modality, or $s$ with an $\mathsf{X}$-modality, then their alternation depths are at most $m-2$. 

  We finally turn to the case $m = 2$, and let $\mathsf{X}_s\mathsf{X}_{t} \in R^X_{1,2(n-k)}$ with $|s|,|t| \leq n-k$. Clearly, $\mathsf{X}_s\mathsf{X}_t$ being defined on a word $w$ is equivalent to $\mathsf{X}_s$ and $\mathsf{Y}_{\bar{t}}$ being defined and $\mathsf{X}_s(w) < \mathsf{Y}_{\bar{t}}(w)$ holding. Suppose that $\mathsf{X}_s(v) < \mathsf{Y}_{\bar{t}}(v)$, then since $u \leq^{\YX}_{2,n} u'$, we have
  \begin{equation}\label{eqn:subwordsubranker}
	  \mathsf{Y}_q\mathsf{X}_s(u) < \mathsf{X}_p\mathsf{Y}_{\bar{t}}(u) \Rightarrow \mathsf{Y}_q\mathsf{X}_s(u') < \mathsf{X}_p\mathsf{Y}_{\bar{t}}(u').
  \end{equation}
  In particular, $\mathsf{Y}_q\mathsf{X}_{s'}(u) < \mathsf{X}_p$ for all prefixes $s'$ of $s$. Hence $\mathsf{X}_s$ is defined on $u'$ and analogously, so is $\mathsf{Y}_{\overline{t}}$. Thus (\ref{eqn:subwordsubranker}) in particular implies $\mathsf{X}_s(v') < \mathsf{Y}_{\overline{t}}(v')$ which implies that $\mathsf{X}_{s}\mathsf{X}_t$ is defined on $u$. 
  By a symmetrical argument, the same holds for rankers in $R^Y_{1,2n}$.
\end{proof}

This lemma also has a counterpart for the $\XX$ and $\YY$-levels. Note that in this case, we lose depth but no alternation depth when considering factors.

\begin{lemma}\label{lem:subwordsubrankerXX}
	Let $m \geq 2$ and let $u,u' \in A^*$ such that
	\begin{equation*}
		u = xavby \equiv^{\XX}_{m,n} x'av'by' = u'
	\end{equation*}
	If there exists rankers $\mathsf{X}_{p}$, $\mathsf{X}_q$ with $|p|,|q| \leq k < n$ such that $\mathsf{X}_p(u) = |xa|$, $\mathsf{X}_p(u') = |x'a|$, $\mathsf{X}_q(u) = |xavb|$ and $\mathsf{X}_q(u') = |x'av'b|$, then $v \equiv^{\XX}_{m,n-k} v'$.
	We also allow $xa$ and $x'a$ (resp. $by$ and $by'$) to be empty, in which case we interpret $\mathsf{X}_p$ (resp. $\mathsf{X}_q$) to be the empty word, $|p| = 0$ (resp. $|q| = 0$) and $|\mathsf{X}_p(u)| = |\mathsf{X}_p(v)| = 0$ (resp. $|\mathsf{X}_p(u)| = |u| + 1$, $|\mathsf{X}_q(v)| = |v| + 1$).
\end{lemma}

\begin{proof}
	By symmetry, it is enough to show that a ranker defined on $v$ is also defined on $v'$ and if $r(v) \leq s(v)$, then $r(v') \leq s(v')$. 
	
	Suppose $r \in [\XX_{m,n-k}]$ starts with an $\mathsf{X}$-modality. If $r$ is defined on $v$, then if follows from Lemma \ref{lem:DefinedOnFactor} that $\mathsf{X}_pr$ is defined on $u$ and $\mathsf{X}_p(u) < \mathsf{X}_pr'(u) < \mathsf{X}_q(u)$ for all nonempty prefixes of $r$. Since $u \equiv^{\XX}_{m,n} u'$, we get that $\mathsf{X}_pr$ is defined on $u'$ and $\mathsf{X}_p(u') < \mathsf{X}_pr'(u') < \mathsf{X}_q(u')$. Again by Lemma \ref{lem:DefinedOnFactor}, it follows that $r$ is defined on $v'$. If $r$ starts with a $\mathsf{Y}$-modality, we make the same argument using $\mathsf{X}_qr'$ instead of $\mathsf{X}_pr'$.
	
	Let $(r,s) \in \XX_{m,n-k}$ where both $r$ and $s$ are defined on $v$ (and hence on $v'$) such that $r(v) \leq s(v)$. Suppose both rankers start with $\mathsf{X}$-modalities. Then
  \begin{alignat*}{2}
	  r(v) < s(v)
	  \quad \Rightarrow \quad & \mathsf{X}_pr(u) < \mathsf{X}_ps(u)
    &\qquad\quad&
    \\ \Rightarrow \quad & \mathsf{X}_pr(u') < \mathsf{X}_ps(u')
    &\qquad\quad&\text{Since $u \equiv^{\XX}_{m,n} u'$}
    \\ \Rightarrow \quad & r(v') < s(v')
    &\qquad\quad&
  \end{alignat*}
	The other cases are similar.
\end{proof}

We now restate the known correspondences between rankers and  the quantifier alternation hierarchy in our framework.  The following characterization of the $\FO^2_m$ levels is due to Weis and Immerman (\itref{aaa:FOcomparisons} and \itref{bbb:FOcomparisons}  \cite{WeisImmerman2009lmcs}) and Kuf\-leitner and Weil (\itref{aaa:FOcomparisons} and \itref{ccc:FOcomparisons} \cite{kufleitnerweil2012csl}).

\begin{proposition}\label{prp:FOcomparisons}
	Given a language $L$, the following are equivalent:
	\begin{enumerate}[(i)]
		\item $L$ is definable in $\FO^2_{m}$,\label{aaa:FOcomparisons}
		\item $L$ is definable by $\XY_{m,n} \cup \YX_{m,n}$ for some $n$,\label{bbb:FOcomparisons}
		\item the syntactic morphism of $L$ is in $\mathbf{R}_{m+1} \cap \mathbf{L}_{m+1}$.\label{ccc:FOcomparisons}
	\end{enumerate}
\end{proposition}

Similarly, we have the following characterization of the $\Sigma^2_m$ levels, due to Fleischer et al.\ (\itref{aaa:sigmaeqv} and \itref{ccc:sigmaeqv} \cite{fleischerkufleitnerlauser2014csr}) and Lauser (\itref{aaa:sigmaeqv} and \itref{bbb:sigmaeqv} \cite[Thm. 11.3]{Lauser2014thesis}). One easily gets the symmetric characterization of the $\Pi^2_m$-levels.

\begin{proposition}\label{prp:XYcomparisons}
	Given a language $L$, the following are equivalent:
	\begin{enumerate}[(i)]
		\item $L$ is definable in $\Sigma^2_{m}$,\label{aaa:sigmaeqv}
		\item $L$ is definable by $\XY_{m,n}$ for odd $m$ and $\YX_{m,n}$ for even $m$ for some $n$,\footnote{The relation in \cite{Lauser2014thesis} has only one-sided inclusion of definedness of rankers with the maximum number of alternations for all $m$ as an explicit assumption. However, for $m \geq 2$, two sided inclusion follow implicitly.}\label{bbb:sigmaeqv}
		\item the syntactic monoid of $L$ is in $\mathbf{Si}_m$.\label{ccc:sigmaeqv}
	\end{enumerate}
\end{proposition}

\begin{proof}
	We only need to show that the ranker comparisons defined by Lauser is equivalent to ours up to the depth of the rankers. Recall Definition 11.2 in \cite{Lauser2014thesis}, defining $\leq^R_{m,n}$ as follows.
	If $m = 0$ or $n = 0$, then $u \leq^R_{m,n} v$ for all $u,v \in A^*$. Otherwise, $u \leq^{R}_{m,n} v$ if $v \leq^R_{m-1,n} u$ and:\footnote{Note that the direction of $\leq^R_{m,n}$ is reversed compared to Definition \ref{def:comparisonssets}. In the former, if $u \leq^{\mathscr{C}} v$, then $r(u) \leq s(u)$ implies $r(v) \leq s(v)$ for $r,s \in [\mathscr{C}]$, while if $u \leq^{R}_{m,n} v$, then $r(v) \leq s(v)$ implies $r(u) \leq s(u)$ for the relevant $r$ and $s$.}
	\begin{enumerate}[\itshape(i)]
		\item for $r \in R_m$ such that $|r| \leq n$, if $r(v)$ is defined then $r(u)$ is defined,\label{aaa:XYcomparisons}
		\item if $r \in R^X_m$, $s \in R^X_{m-1}$ such that $|r| \leq n$, $|s| \leq n - 1$, then if $m$ is odd, then $r(v) \leq s(v)$ implies $r(u) \leq s(u)$ and $r(v) < s(v)$ implies $r(u) < s(u)$ and if $m$ is even, then $r(v) \geq s(v)$ then $r(u) \geq s(u)$ and if $r(v) > s(v)$, then $r(u) > s(u)$,
		\item if $r \in R^Y_m$, $s \in R^Y_{m-1}$ such that $|r| \leq n$, $|s| \leq n - 1$, then if $m$ is odd, then $r(v) \geq s(v)$ implies $r(u) \geq s(u)$ and $r(v) > s(v)$ implies $r(u) > s(u)$ and if $m$ is even, then $r(v) \leq s(v)$ then $r(u) \leq s(u)$ and if $r(v) < s(v)$, then $r(u) < s(u)$,
		\item if $r \in R^Y_m$, $s \in R^Y_{m}$ such that $|r| + |s| < 2n$, then if $m$ is odd, then $r(v) \leq s(v)$ implies $r(u) \leq s(u)$ and $r(v) < s(v)$ implies $r(u) < s(u)$ and if $m$ is even, then $r(v) \geq s(v)$ then $r(u) \geq s(u)$ and if $r(v) > s(v)$, then $r(u) > s(u)$.
	\end{enumerate}
	We claim that for $m \geq 2$, we can without loss of generality strengthen \itref{aaa:XYcomparisons} to read
	\begin{enumerate}[\itshape(i')]
		\item for $r \in R_m$ such that $|r| \leq n$, we have $r(v)$ defined if and only if $r(u)$ is defined,
	\end{enumerate}
	in which case the result is obvious.

	Without loss of generality, assume that $r = r'\mathsf{X}_p\mathsf{Y}_q$ where $r'$ has alternation depth $m - 2$ or is empty. Since $u \leq^R_{m,n} v$ implies $v \leq^R_{m-1,n} u$ by definition, we have that $r'\mathsf{X}_p$ and $\mathsf{X}_{\overline{q}}$ are both defined on $v$. If $r(v)$ is not defined, we must have $r'\mathsf{X}_p(v) < \mathsf{X}_{\overline{q}}(v)$ which implies $r'(v) < \mathsf{X}_{\overline{q}}\mathsf{Y}_{\overline{p}}(v)$ (or if $r'$ is empty, it implies that $\mathsf{X}_{\overline{q}}\mathsf{Y}_{\overline{p}}(v)$ is defined). It follows that $r'(u) < \mathsf{X}_{\overline{q}}\mathsf{Y}_{\overline{p}}(u)$ which contradicts $r(u)$ being defined.
\end{proof}

The corners of the Trotter-Weil hierarchy has a ranker characterization in terms of condensed rankers \cite{KufleitnerWeil2012lmcs}. We reformulate this result in terms of ranker comparisons. The ranker comparison characterization of the join levels then follow directly. We also use Proposition \ref{prp:rvlandsigmacappi} to relate the join levels to the intersection levels of the quantifier alternation hierarchy.

\begin{proposition}\label{prp:RRcomparisons}
	Let $m \geq 1$. Then $L$ is definable by $\XX_{m,n}$ (resp. $\YY_{m,n}$) for some $n$ if and only if its syntactic monoid $M$ is in $\mathbf{R}_{m+1}$ (resp. $\mathbf{L}_{m+1}$). Furthermore, the following are equivalent:
	\begin{enumerate}[(i)]
		\item $L$ is definable by $\XX_{m,n} \cup \YY_{m,n}$ for some $n$, \label{aaa:RRcomparisons}
		\item the syntactic monoid of $L$ is in $\mathbf{R}_{m+1} \vee \mathbf{L}_{m+1}$, \label{bbb:RRcomparisons}
		\item the syntactic monoid of $L$ is in $\Pivar_{m} \cap \mathbf{Si}_{m}$, \label{ccc:RRcomparisons}
		\item $L$ is definable in $\Sigma_{m}^2$ and in $\Pi_{m}^2$. \label{ddd:RRcomparisons}
	\end{enumerate}
\end{proposition}

\begin{proof}
	Let $\mu: A^* \to M$ be the syntactic morphism of $L$. We use induction on $m$. We consider the case $\mathbf{R}_{m+1}$ with $\mathbf{L}_{m+1}$ being symmetric. Suppose first that $L$ is definable by $\XX_{m,n}$, and $M$ is its syntactic monoid. Since $M$ is a quotient of $N^{\XX}_{m,n}$, it is enough to show $N^{\XX}_{m,n} \in \mathbf{R}_{m+1}$. Let $\nu: A^* \to N^{\XX}_{m,n}$ and let $\pi: N^{\XX}_{m,n} \to \faktor{N^{\XX}_{m,n}}{\sim_{\mathbf{K}}}$ be the natural projection. Let $u,v \in A^*$ be such that $u \equiv^{\YY}_{m-1,n} v$ if $m \geq 2$, or $\alp(u) = \alp(v)$ for $m = 1$. We show that $\mu(u) \sim_{\mathbf{K}} \mu(v)$.

	Let $e$ be idempotent such that $e \Jeq e\mu(u)$. Then there exists $\hat{e} \in \mu^{-1}(e)$ with $\alp(u) = \alp(v) \subseteq \alp(\hat{e})$. Since $e$ is idempotent, we have $\hat{e}^{2n} \in \mu^{-1}(e)$. We want to show $\hat{e}^{2n}u \equiv^{\XX}_{m,n} \hat{e}^{2n}v$.
	
	We consider rankers $r$ and $s$. Suppose first that both start with an $\mathsf{X}$-modality. In particular, for $m = 1$ this is the only possibility. We note that by the length of $\hat{e}^n$ and the alphabetic conditions, $r$ and $s$ are defined on $\hat{e}^{2n}u$ if and only if they are defined on $\hat{e}^{n}$. The same holds for $\hat{e}^{2n}v$. In particular, they are defined on $\hat{e}^{2n}u$ if and only if $\hat{e}^{2n}v$. Furthermore,
	\begin{equation*}
		r(\hat{e}^{2n}u) \leq s(\hat{e}^{2n}v) \Leftrightarrow r(\hat{e}^{n})  \leq s(\hat{e}^{n}) \Leftrightarrow r(\hat{e}^{2n}v) \leq s(\hat{e}^{2n}v)
	\end{equation*}
	and analogously for the strict relation.

	Next, suppose $r$ and $s$ are rankers which start with a $\mathsf{Y}$-modality. Then $(r,s) \in \YY_{m-1,n}$. Since $u \equiv^{\YY}_{m-1,n} v$ it follows by stability that $\hat{e}^{2n}u \equiv^{\YY}_{m-1,n} \hat{e}^{2n}v$. In particular, the same such rankers are defined on $u$ and $v$. 

	Finally, suppose $r$ starts with an $\mathsf{X}$-modality, and $s$ starts with a $\mathsf{Y}$-modality. It again follows from the length of $\hat{e}^n$ and the alphabetic conditions that $r(\hat{e}^{2n}u) < s(\hat{e}^{2n}u)$ if and only if both are defined on $\hat{e}^{2n}u$ and similarly for $\hat{e}^{2n}v$. For the same reason, none of the words can satisfy $r \leq s$ if $r$ starts with a $\mathsf{Y}$-modality and $s$ starts with an $\mathsf{X}$-modality.

	Since this covers all possible cases, we get $\hat{e}^{2n}u \equiv^{\XX}_{m,n} \hat{e}^{2n}v$. This implies $e\nu(u) = e\nu(v)$ and since $e$ was arbitrary such that $e \Jeq e\nu(u)$, it follows that $\nu(u) \sim_{\mathbf{K}} \nu(v)$. For $m = 1$, this implies that $\pi \circ \nu$ factors through $J_A$ and for for $m \geq 2$ that $\pi \circ \nu$ factors through $N^{\YY}_{m-1,n}$. In either case, $\faktor{N^{\XX}_{m,n}}{\sim_{\mathbf{K}}} \in \mathbf{L}_{m}$ (for $m = 1$ since $J_A \in \mathbf{L}_1$ and for $m \geq 2$ since $N^{\YY}_{m,n} \in \mathbf{L}_{m+1}$ by induction). Thus $N^{\XX}_{m,n} \in \mathbf{R}_{m+1}$.
	
	For the other direction, suppose $\pi: M \to \faktor{M}{\sim_{\mathbf{K}}} \in \mathbf{L}_{m}$ is the canonical projection. We will use induction to show that $u \equiv^{\XX}_{m,m|M|+1} v$ implies $\mu(u) = \mu(v)$.
	Suppose that $u \equiv^{\XX}_{m,m|M|+1} v$ and let $u = u_0 a_1 u_1 \cdots a_n u_n$ be the \greenR-factorization of $u$. We factor $v = v_0 a_1 v_1 \cdots a_n v_n$ where $a_i \notin v_{i-1}$. Such a factorization exists since $u \equiv^{\XX}_{m,m|M|+1} v$. Furthermore, we have
  \begin{alignat*}{3}
	  b \in \alp(u_i)  & \Leftrightarrow \mathsf{X}_{a_1 \dots a_i}(u) <  \mathsf{X}_{a_1 \dots a_ib}(u) < \mathsf{X}_{a_1 \dots a_ia_{i+1}}(u) & \qquad\qquad & \text{}
	  \\& \Leftrightarrow \mathsf{X}_{a_1 \dots a_i}(v) <  \mathsf{X}_{a_1 \dots a_ib}(v) < \mathsf{X}_{a_1 \dots a_ia_{i+1}}(v) & \qquad\qquad &
	  \\& \Leftrightarrow b \in \alp(v_i). & \qquad\qquad & 
  \end{alignat*}
  Thus, $\alp(u_i) = \alp(v_i)$ for all $i$ (the argument for $i = 0$ and $i = n$ is similar, but uses definedness and one comparison instead of two comparisons). Furthermore, for $m \geq 2$, Lemma \ref{lem:subwordsubrankerXX} shows that $u_i \equiv^{\XX}_{m,(m-1)|M|+1} v_i$ which in particular implies $u_i \equiv^{\YY}_{m-1,(m-1)|M|+1} v_i$. By induction, this implies $\mu(u_i) \sim_{\mathbf{K}} \mu(v_i)$ for all $m$. Let $x_i$ be such that $\mu(u_0 a_1 \dots a_i u_i x_i) = \mu(u_0 a_1 \dots a_i)$. We get
	\begin{align*}
		\mu(u_0 a_1 u_1 \ldots a_n u_n) & = \mu(v_0 a_1 u_1 \ldots a_n (u_nx_n)^{\omega} u_n) 
		\\ & = \mu(u_0 a_1 u_1 \ldots a_n (u_nx_n)^{\omega} v_n) 
		\\ & = \mu(u_0 a_1 u_1 \ldots a_n v_n) 
		\\ & \;\, \vdots
		\\ & = \mu(v_0 a_1 v_1 \ldots a_n v_n).
	\end{align*}
	Thus $\mu$ factors through $N^{\XX}_{m,n+|M|}$ which gives the desired result.

	We now turn to the join levels. The language $L$ has its syntactic monoid in $\mathbf{R}_{m+1} \vee \mathbf{L}_{m+1}$ if it is a divisor of some $M_1 \times \cdots \times M_n$ where $M_i \in \mathbf{R}_{m+1} \cup \mathbf{L}_{m+1}$. In particular, the language is recognized by some $\mu: A^* \to M_1 \times \cdots \times M_n$. Let $\mu_i = \pi_i \circ \mu$ where $\pi_i$ is the projection on the $i^{\text{th}}$ monoid. Let $u = (u_1,\dots,u_n) \in M_1 \times \dots \times M_n$. By the previous part, $\mu^{-1}_{i}(u_i)$ is defineable by $\XX_{m,n} \cup \YY_{m,n}$ for some $n$. Since $\mu^{-1}(u) = \bigcap_{i} \mu^{-1}_i(u_i)$, so is $\mu^{-1}(u)$. Since it is the union of such sets, it follows that $L$ is definable by $\XX_{m,n} \cup \YY_{m,n}$.
	
	For the other direction, suppose $L$ is definable in $\XX_{m,n} \cup \YY_{m,n}$. Then $L$ is a Boolean combination of languages $L_i$ definable in $\XX_m$ or $\YY_m$. Let $\mu_i: A^* \to M_i$ be the syntactic morphism of $L_i$, and define $\mu: A^* \to M_1 \times \dots \times M_n$ by $\mu(u) = \left( \mu_1(u), \dots, \mu_n(u) \right)$. It is clear that $L$ is the preimage of a union of elements in $M_1 \times \dots \times M_n$, and it follows that the syntactic monoid of $L$ is in $\mathbf{R}_{m+1} \vee \mathbf{L}_{m+1}$.
	
	Finally, the equivalence between \itref{ddd:RRcomparisons} and \itref{bbb:RRcomparisons} is Proposition \ref{prp:rvlandsigmacappi} and the equivalence between \itref{ccc:RRcomparisons} and \itref{ddd:RRcomparisons} is an obvious consequence of Proposition \ref{prp:XYcomparisons}.
\end{proof}

Taken together, these three propositions gives us a new way of considering the Trotter-Weil hierarchy and the quantifier alternation hierarchy together, as a ranker comparison hierarchy; see Figure \ref{fig:hierarchies}.

\section{Saturations for Fragments of $\FO^2$}
\label{sec:decidingconelikes}

In this section, we present computable closed sets for all levels of the Trotter-Weil and quantifier alternation hierarchies, in other words for all levels of the ranker comparison hierarchy. We also state our main results: that these sets agree with the corresponding sets of pointlikes. The proof thereof is the subject of the subsequent sections.

The sets presented below relies on the monoids having content morphisms (intuitively on the monoid elements having a fixed alphabet). This is not true for all monoids; consider for example $M = \left\{ 1,a \right\}$ with $aa = 1$. However, it is always possible to \emph{alphabetize} a monoid by explicitly distinguishing elements with different alphabets. If $M$ is a monoid with a generating set $A$, then the submonoid of $M \times J_A$ generated by $(a,\left\{ a \right\})_{a \in A}$ has a content morphism. It also has a surjective morphism onto $M$. The following Lemma shows that this is enough to deduce the pointlikes of $M$.\footnote{A variant of the Lemma can be found in \cite{AlmeidaCarlosZeitoun2008jpaa}.}

\begin{lemma}\label{lem:alphabetsdoesntmatter}
	Let $M, M'$ be finite monoids and suppose that there is a surjective morphism $\pi : M' \to M$. Then $P \subseteq M$ (resp. $(s,S) \in M \times 2^M$) is pointlike (resp. conelike) with respect to a variety $\mathbf{V}$ if and only if there exists a pointlike $P' \subseteq M'$ (resp. conelike $(s',S') \in M' \times 2^{M'}$) with respect to $\mathbf{V}$ such that $P \subseteq \pi(P')$ (resp. $S \subseteq \pi(S')$, $s = \pi(s')$).
\end{lemma}

\begin{proof}
	We give the proof for pointlikes; the proof for conelikes is analogous.
	Let $\tau: M \to N \in \mathbf{V}$ be a relational morphism. Then $\tau' = \tau \circ \pi^{-1}$ is also a relational morphism. If $P'$ is pointlike with respect to $\mathbf{V}$, then in particular it is pointlike with respect to $\tau'$. It follows that $\pi(P')$ and thus $P$ must be pointlike with respect to $\tau$. Since $\tau$ was arbitrary, this is true for $\mathbf{V}$.

	For the other direction, let $P \subseteq M$ and let $P_1,\dots,P_n$ be the collection of all subsets of $M'$ satisfying $P \subseteq \pi(P_i)$. If $P_i$ is not pointlike with respect to $\mathbf{V}$, then there exists $\tau_i: M' \to N_i \in \mathbf{V}$ such that $P_i$ is not pointlike with respect to $\tau_i$. Suppose that no $P_i$ is pointlike. We define $\tau: M' \to N_1 \times \dots \times N_n$ by
	\begin{equation*}
		\tau(s) = \left\{ (x_1,\dots,x_n) \mid x_i \in \tau_i(s) \right\}.
	\end{equation*}
	It is straightforward to check that $\tau$ is a relational morphism. We let $\tau': \tau \circ \pi^{-1}$. For contradiction, suppose that $P$ is pointlike with respect to $\mathbf{V}$, then in particular $P$ is pointlike with respect to $\tau'$. Thus, there exists $(x_1,\dots,x_n)$ such that $P \subseteq \tau'^{-1}(x_1,\dots,x_n)$. Let $U = \tau^{-1}(x_1,\dots,x_n)$. It is clear that $P \subseteq \pi(U)$, and thus $U = P_i$ for some $i$. Since $(x_1,\dots,x_i,\dots,x_n) \in \tau(s)$ for all $s \in P_i$, we must have $x_i \in \tau_i(s)$ for all $s \in P_i$, a contradiction.
\end{proof}

We now introduce the relevant closed sets. Note that for our purposes, $\mathbf{R}_1 = \mathbf{L}_1 = \mathbf{J}_1$. We first give the sets for the corners of the Trotter-Weil hierarchy. These are important building blocks for the other sets.

\begin{definition}\label{def:RLSaturations}
	Let $M$ be a monoid with a content morphism $\alpha$. We define:
	\begin{itemize}
		\item $ \sat{\mathbf{J}_1}{M} = \satR{1}{M} = \satL{1}{M} = \left\{ S \subseteq M \mid \alpha(s) = \alpha(t) \text{ for all $s,t \in S$} \right\}$,
		\item for $m \geq 2$, $\satR{m}{M}$ is the smallest closed set of $M$ such that if $Z \in \satL{m-1}{M}$, $U \in \satR{m}{M}$, $\alpha(Z) \leq \alpha(U)$ and $U$ is idempotent in $2^M$, then $UZ \in \satR{m}{M}$
		\item for $m \geq 2$, $\satL{m}{M}$ is the smallest closed set of $M$ such that if $Z \in \satR{m-1}{M}$, $V \in \satL{m}{M}$, $\alpha(Z) \leq \alpha(V)$ and $V$ is idempotent in $2^M$, then $ZV \in \satL{m}{M}$
	\end{itemize}
\end{definition}

The definition inductively ensures that for any $W$ in any of the introduced sets, we have $\alpha(w) = \alpha(w')$ for all $w,w' \in W$. Thus, $\alpha(W)$ is a well defined element of $J_A$, making the comparisons $\alpha(Z) \leq \alpha(U)$ and $\alpha(Z) \leq \alpha(V)$ meaningful.

The other closed sets build on so-called $\mathbf{RL}_m$-factors. If one think of the elements of a monoid as the languages they represent, one can think of $\mathbf{RL}_m$-factors as collections of languages which can not be distinguished from any side using rankers of alternation depth at most $m$, while containing words of arbitrary length. 

\begin{definition}\label{def:RLmFactors}
	Let $M$ be a monoid with a content morphism $\alpha$. Let $S, E \in \satR{m}M$, $T, F \in \satL{m}M$ with $\alpha(S),\alpha(T) \leq \alpha(E) = \alpha(F)$ and $E$, $F$ idempotent in $2^M$. Let $W$ be such that $\alpha(w) \leq \alpha(E)$ for all $w \in W$. Then $SEWFT$ is an \emph{$\mathbf{RL}_m$-factor}.
\end{definition}

Since $\satR{m}M$ and $\satL{m}M$ can be constructed for each $m$, the $\mathbf{RL}_m$-factors can also be effectively constructed. Note that the alphabet of an $\mathbf{RL}_m$-factor is well defined. Using these factors, we construct the following sets.

\begin{definition}\label{def:SaturationDefinitions}
	Let $M$ be a monoid. Then 
	\begin{itemize}
		\item $\psatS{1}{M}$ is the smallest closed set for which $(1,S) \in M$ for all $S \subseteq M$.
		\item $\psatP{1}{M}$ is the smallest closed set for which $(s,\left\{ 1,s \right\}) \in M$ for all $s \in M$.
	\end{itemize}
	Suppose further that $M$ has a content morphism $\alpha$, then for $m \geq 2$:
	\begin{itemize}
		\item $\sat{\mathbf{J}}{M}$, is the smallest closed set such that 
	\begin{equation*}
		XEY \in \sat{\mathbf{J}}{M}
	\end{equation*}	
	for all $X,Y,E \in 2^M$ where $E$ is idempotent, $\alpha(s) = \alpha(t)$ for all $s,t \in E$, and $\alpha(w) \subseteq \alpha(E)$ for all $w \in X \cup Y$.
		\item  $\satFO{m+1}{M}$ is the smallest closed set which for all $n$ contain the product
\begin{equation*}
	U_1 V_1 U_2 \dots V_{n-1} U_n
\end{equation*}
where every $U_i$ is an $\mathbf{RL}_m$-factor while $V_i \in \satFO{m}M$ (or $V_i  \in \sat{\mathbf{J}}M$ for $m = 2$), and $\alpha(v_i') \leq \alpha(U_i), \alpha(U_{i+1})$ for all $v'_i \in V_i$,
		\item  $\satSP{m}{M}$ is the smallest closed set which for all $n$ contain the product
\begin{equation*}
	U_1 V_1 U_2 \dots V_{n-1} U_n
\end{equation*}
where every $U_i$ is an $\mathbf{RL}_m$-factor while $V_i \in \satSP{m-1}M$ (or $V_i \in 2^M$ for $m = 2$), and $\alpha(v_i') \leq \alpha(U_i), \alpha(U_{i+1})$ for all $v'_i \in V_i$,
		\item  $\psatS{m}{M}$ is the smallest closed set which for all $n$ contain the product
\begin{equation*}
	(u_1,U_1) (v_1,V_1) (u_2,U_2) \dots (v_{n-1},V_{n-1}) (u_n,U_n)
\end{equation*}
where for all $i$, we have $u_i \in U_i$ and $U_i$ is an $\mathbf{RL}_m$-factor while $(v_i,V_i) \in \psatS{m-1}M$, and $\alpha(v_i') \leq \alpha(U_i), \alpha(U_{i+1})$ for all $v'_i \in V_i$,
		\item  $\psatP{m}{M}$ is the smallest closed set which for all $n$ contain the product
\begin{equation*}
	(u_1,U_1) (v_1,V_1) (u_2,U_2) \dots (v_{n-1},V_{n-1}) (u_n,U_n)
\end{equation*}
where for all $i$, we have $u_i \in U_i$ and $U_i$ is an $\mathbf{RL}_m$-factor while $(v_i,V_i) \in \psatP{m-1}M$, and $\alpha(v_i') \leq \alpha(U_i), \alpha(U_{i+1})$ for all $v'_i \in V_i$,
	\end{itemize}
\end{definition}

We now state our main theorem. Apart from giving the pointlikes of the different levels, it also provides \emph{separators}. These are monoids with relational morphisms which are optimal in $\mathbf{V}$ for separating the elements of $M$. In other words, the relational morphisms $\tau$ satisfy $\pl{\tau}M = \pl{\mathbf{V}}M$. The theorem states only the monoids explicitly; the relational morphisms are the natural relational morphisms, obtained by mapping every element in $M$ to their preimage in $A^*$ and projecting onto the relevant monoids.

\begin{theorem}\label{thm:Pointlikes}
	Let $M$ be a finite monoid, and let $n = \ceil{R/2} - 1$ where $R$ is the Ramsey number of $M$. Then
	\begin{enumerate}[(i)]
		\item $\coneS{1}{M} = \psatS{1}M$ with separator $N^{\XY}_{1,n}$,\label{aaa:ConelikesForSiAndPiLevels}
		\item $\coneP{1}{M} = \psatP{1}M$ with separator $N^{\YX}_{1,n}$,\label{bbb:ConelikesForSiAndPiLevels}
	\setcounter{resumeCounter}{\value{enumi}}
	\end{enumerate}
	Furthermore, suppose $M$ has a content morphism $\alpha: M \to J_A$, and let $n = (m + |A|)(R-1)$ and $n' = (m-1+3|A|)(R-1) + |A|$ where $R$ is the Ramsey number of $2^M$.
	\begin{enumerate}[(i)]
	\setcounter{enumi}{\value{resumeCounter}}
		\item $\pl{\mathbf{J}_1}{M} = \sat{\mathbf{J}_1}M$ with the separator $J_A$,\label{aaa:PointlikesForRandLLevels}
		\item $\plR{m}{M} = \satR{m}M$ with the separator $N^{\XX}_{m,n}$,\label{bbb:PointlikesForRandLLevels}
		\item $\plL{m}{M} = \satL{m}M$ with the separator $N^{\YY}_{m,n}$,\label{ccc:PointlikesForRandLLevels}
\item  $\pl{\mathbf{J}}{M} = \sat{\mathbf{J}}M$ with separator $N^{\XY \cup \YX}_{1,|A|R+R-1}$,\footnote{See.~\cite{AlmeidaCarlosZeitoun2008jpaa}. We reprove it in order to get a separator defined using rankers.}\label{bbc:ConelikesForSiAndPiLevels}
		\item  $\plFO{m}{M} = \satFO{m+1}M$ with separator $N^{\XX \cup \YY}_{m,n'}$,\label{ccc:ConelikesForSiAndPiLevels}
		\item  $\plSP{m}{M} = \satSP{m}M$ with separator $N^{\XX \cup \YY}_{m,n'}$,\label{ccd:ConelikesForSiAndPiLevels}
		\item  $\coneS{m}{M} = \psatS{m}M$ with separator $N^{\XY}_{m,n'}$ for odd $m$ and $N^{\YX}_{m,n'}$ for even $m$,\label{ddd:ConelikesForSiAndPiLevels}
		\item  $\coneP{m}{M} = \psatP{m}M$ with separator $N^{\YX}_{m,n'}$ for odd $m$ and $N^{\XY}_{m,n'}$ for even $m$,\label{eee:ConelikesForSiAndPiLevels}
	\end{enumerate}
\end{theorem}



The following is an immediate corollary, given Proposition \ref{prp:conelikesandpointedcoverrelation}.

\begin{corollary}
	The covering problem has a solution for all language varieties associated with the levels of the quantifier alternation hierarchy. 
	In particular, this implies solutions to the separation problems for all of these varieties.
\end{corollary}

\begin{example}\label{exm:SaturationExample}
	Let $M$ be the syntactic monoid of $(ab)^+$; see Figure \ref{fig:exampleM}. Note that this monoid does not have a content morphism since (e.g.)~$a$ and $aba$ maps to the same element in $M$. Thus, we instead consider the alphabetized monoid $M' = M \times J_{ \left\{ a,b \right\}}$ shown in Figure \ref{fig:exampleMprime}.
	\begin{figure}
		\centering
		\begin{subfigure}{0.5\textwidth}
			\centering
		\begin{tikzpicture}
			\draw (0,0) rectangle (1,1) node [pos=.5] {$1$};
			\draw (-.5,-1.5) rectangle ++(1,1) node [pos=.5] {$a$};
			\draw (.5,-1.5) rectangle ++(1,1) node [pos=.5] {$ab$};
			\draw (-.5,-2.5) rectangle ++(1,1) node [pos=.5] {$ba$};
			\draw (.5,-2.5) rectangle ++(1,1) node [pos=.5] {$b$};
			\draw (0,-4) rectangle ++(1,1) node [pos=.5] {$aa$};
		\end{tikzpicture}
		\caption{The monoid $M$}\label{fig:exampleM}
		\end{subfigure}\begin{subfigure}{0.5\textwidth}
			\centering
		\begin{tikzpicture}
			\draw (0,1.5) rectangle ++(1,1) node [pos=.5] {$1$};
			\draw (-0.75,0) rectangle ++(1,1) node [pos=.5] {$a$};
			\draw (0.75,0) rectangle ++(1,1) node [pos=.5] {$b$};
			\draw (-.5,-1.5) rectangle ++(1,1) node [pos=.5] {$aba$};
			\draw (.5,-1.5) rectangle ++(1,1) node [pos=.5] {$ab$};
			\draw (-.5,-2.5) rectangle ++(1,1) node [pos=.5] {$ba$};
			\draw (.5,-2.5) rectangle ++(1,1) node [pos=.5] {$bab$};
			\draw (-2,-2) rectangle ++(1,1) node [pos=.5] {$aa$};
			\draw (2,-2) rectangle ++(1,1) node [pos=.5] {$bb$};
			\draw (0,-4) rectangle ++(1,1) node [pos=.5] {$aab$};
		\end{tikzpicture}
		\caption{The monoid $M' = M \times J_{\left\{ a,b \right\}}$}\label{sub:aAstarbwithgroup}\label{fig:exampleMprime}
		\end{subfigure}
		\caption{}\label{fig:example}
	\end{figure}

	In Example \ref{exm:PointlikeExample}, we note that $(ab,\left\{ aba,bab \right\})$ is conelike with respect to $\mathbf{J}^+ = \Si_1$. To see that this follows from Theorem \ref{thm:Pointlikes}, we show that $(ab,\left\{ aba,bab \right\}) \in \psatS{1}{M'}$. Indeed, we have $(1,\left\{ 1,a \right\}), (1,\left\{ 1,b \right\}) \in \psatS1{M'}$ and thus by closure under multiplication we get $(1,\left\{ 1,b \right\})(ab,\left\{ ab \right\})(1,\left\{ 1,a \right\}) = (ab,\left\{ ab,bab,aba,ba \right\}) \in \psat1{M'}$. It follows that $(ab,\left\{ bab,aba \right\}) \in \psat1{M'}$ by closure under subsets in the second entry.

	Next, we consider the pointlikes with respect to $\mathbf{R}_2$ and $\mathbf{L}_2$, which in particular helps us to calculate the $\mathbf{RL}_m$-factors. We have $\left\{ aba,ab,ba,bab,aab \right\} \in \sat{\mathbf{J}_1}{M'}$ and, for example, $\left\{ ab \right\} \in \satR1{M'}$. Thus $\left\{ ab \right\}\left\{ aba,ab,ba,bab,aab \right\} = \left\{ aba,ab,aab \right\} \in \satR1{M'}$. Similarly, we get $\left\{ ba,bab,aab \right\} \in \satR1{M'}$. The intuition here is that $\mathbf{R}_2$ can distinguish the possible beginnings of words in the respective languages, but no other details. Analogously, $\mathbf{L}_2$ can distinguish the possible endings, and we get $\left\{ ab,bab,aab \right\}, \left\{ aba, ba, aab\right\} \in \satL1{M'}$. 

	Following this intuition, one can consider the $\mathbf{RL}_1$-factors to be the sets where we can distinguish both the beginning and the end. Indeed, calculating the $\mathbf{RL}_1$-factors yields (apart from the singletons) the sets $\left\{ aba,aab \right\}, \left\{ ab, aab \right\}, \left\{ ba,aab \right\}$ and $\left\{ bab,aab \right\}$. A straightforward calculation shows that the $\mathbf{RL}_m$ factors are exactly the $\mathbf{RL}_1$-factors independent of $m$.

	Since the products in Definition \ref{def:SaturationDefinitions} are also valid for $n = 1$, these $\mathbf{RL}_m$-factors are themselves pointlike for all levels above $\mathbf{R}_2 \vee \mathbf{L}_2$. Furthermore, we note that
	\begin{equation*}
		\left\{ aba,aab \right\}M'\left\{ ab,aab \right\} = \left\{ ab,aab \right\}
	\end{equation*}
	and similarly for the other combinations. Thus the $\mathbf{RL}_m$-factors are the only pointlikes for all levels above $\mathbf{R}_m \vee \mathbf{L}_m$.

	Using Lemma \ref{lem:alphabetsdoesntmatter}, we see that there is a surjective morphism $\pi: M' \to M$ such that (e.g.)~$\left\{ aba,aab \right\} \subseteq  M'$ maps to $\left\{ a,aab \right\} \subseteq M$. Thus for each level above $\mathbf{R}_2 \vee \mathbf{L}_2$, the pointlikes of $M$ are the singletons and $\left\{ a, aa\right\}$, $\left\{ ab, aa\right\}$, $\left\{ b, aa\right\}$ and $\left\{ ba, aa\right\}$.
\end{example}

\section{The Cases $\Si_1$ and $\Pivar_1$}
\label{sec:BaseCases}

The closed sets for $\Si_1$ and $\Pivar_1$ do not rely on alphabetic properties, and the proof that they coincide with the pointlikes uses techniques different from the other levels. Thus we devote this section to these instances. The inclusions of the sets in the conelikes are trivial.

\begin{lemma}\label{lem:satimpliescone}
	Let $M$ be a monoid. We have $\psatS{1}{M} \subseteq \coneS{1}{M}$.
\end{lemma}

\begin{proof}
	It is clear that the closure properties for closed sets also hold for conelikes. Thus we only need to show that $(1,S)$ is conelike for any $S \subseteq M$. This follows since $\uparrow 1 = N$ for any $N \in \mathbf{J}^+$.
\end{proof}

\begin{lemma}\label{lem:satimpliesconepi}
	Let $M$ be a monoid. We have $\psatP{1}{M} \subseteq \coneP{1}{M}$.
\end{lemma}

\begin{proof}
	Analogously to the previous proof, we need only show that $(s,\left\{ s,1 \right\})$ is conelike for all $s$. For every $N \in \Pivar_1$, and every $x \in N$ we have $1 \in \uparrow x$. Thus $\left\{ 1,s \right\} \subseteq \tau^{-1}(\uparrow x)$ for each $x \in \tau(s)$ giving the desired result.
\end{proof}

For the other direction, the idea is to find minimal representatives for each element in the monoid, and use knowledge about these to separate the languages. We call these representatives minors.

\begin{definition}
	Let $\mu: A^* \to M$ be a morphism. Given $u \in A^*$, we say that $v \in A^*$ is a \emph{$\mu$-minor} of $u$ if $v$ is a subword of $u$, $\mu(u) = \mu(v)$ and $v$ does not have any strict subwords $w$ such that $\mu(w) = \mu(v)$. In other words, $v$ is a subword of $u$ minimal with respect to the subword relation such that $\mu(v) = \mu(u)$.
\end{definition}

\begin{lemma}\label{lem:longwordsinminors}
	Let $\mu: A^* \to M$ be a morphism, and suppose that the Ramsey number of $M$ is $R$. If $v$ is a $\mu$-minor (of some $u$), then $|v| \leq R - 2$.
\end{lemma}

\begin{proof}
	We argue by contradiction. Suppose $v = v_1 \cdots v_n$ is a $\mu$-minor where $n \geq R-1$ and each $v_i$ is nonempty. Let $\mathcal{G} = (V,E)$ be the complete graph with $V = \left\{ 1, \cdots , n+1 \right\}$. The word $v$ induces an $M$-coloring of $\mathcal{G}$ by setting $c(\left\{ i,i' \right\}) = \mu(v_i \cdots v_{i'-1})$. By Theorem \ref{thm:ramseys}, there is a monocrome triangle. Say for instance that $\mu(v_{j} \cdots v_{k-1}) = \mu(v_{k} \cdots v_{\ell-1}) = \mu(v_j \cdots v_{\ell-1})$. We have
	\begin{align*}
		\mu(v) & = \mu(v_1) \ldots \mu(v_{n})
		\\ & = \mu(v_1) \ldots \mu(v_{j-1}) \mu(v_{j} \ldots v_{\ell-1}) \mu(v_{\ell}) \ldots \mu(v_n)
		\\ & = \mu(v_1) \ldots \mu(v_{j-1}) \mu(v_{j} \ldots v_{k-1}) \mu(v_{\ell}) \ldots \mu(v_n)
		\\ & = \mu(v')
	\end{align*}
	where $v' = v_1 \cdots v_{k-1} v_{\ell} \cdots v_n$. Since $v'$ is a strict subword of $v$, we get the desired contradiction.
\end{proof}

\begin{lemma}\label{lem:coneimpliessat}
	Let $M$ be a monoid with generating set $A$, and let $n = \ceil{R/2}-1$ where $R$ is the Ramsey number of $M$. If $\relmorsmall\tau{A^*}M{N^{\XY}_{1,n}}\mu\nu$ is the natural relational morphism, then $\cone{\tau}{M} \subseteq \psatS{1}{M}$.
\end{lemma}

\begin{proof}
	Let $x \in N^{\XY}_{1,n}$ and suppose that $u \in \nu^{-1}(x)$. Let $\tilde{u}  = a_1 \dots a_k$ be a $\mu$-minor of $u$. Since $|\tilde{u}| \leq R - 2$ and $\tilde{u}$ is a subword of $u$, we must have that $\tilde{u}$ is a subword of $v$ for any $v \in \nu^{-1}(\uparrow x)$. Thus we can factor $\nu^{-1}(\uparrow x) = U_0 a_1 U_1 \cdots a_k U_k$. We have 
	\begin{align*}
		(\mu(u), \mu(\nu^{-1}(\uparrow x))) & = (\mu(\tilde{u}), \mu(U_0a_1U_1 \dots a_kU_k))
		\\ & = (1,\mu(U_0))(\mu(a_1),\left\{ \mu(a_1) \right\})(1,\mu(U_1)) \cdots \left( \mu(a_k) \left\{ \mu(a_k) \right\} \right)(1,\mu(U_k))
		\\ & \in \psatS{1}{M}
	\end{align*}
	giving the desired result.
\end{proof}

\begin{lemma}\label{lem:coneimpliessatpi}
	Let $M$ be a monoid with generating set $A$, and let $n = \ceil{R/2}-1$ where $R$ is the Ramsey number of $M$. If $\relmorsmall\tau{A^*}M{N^{\YX}_{1,n}}\mu\nu$ is the natural relational morphism, then $\cone{\tau}{M} \subseteq \psatP{1}{M}$.
\end{lemma}

\begin{proof}
	Note that $N^{\YX}_{1,n}$ is $N^{\XY}_{1,n}$ with the order reversed. In other words, if $u, v \in A^*$, then $\nu(u) \leq \nu(v)$ if and only if every subword of length $2\ceil{R/2}-2$ in $v$ is also in $u$.

	Let $s = \mu(u)$ where $u = a_1 \dots a_n$ for some $a_i \in A^*$. Then 
	\begin{equation*}
		(\mu(a_1),\left\{ 1,\mu(a_1) \right\}) \dots (\mu(a_n), \left\{ 1,\mu(a_n) \right\}) = (s,S) \in \psatP{1}{M}
	\end{equation*}
	where
	\begin{equation*}
		S = \left\{ t \in M \mid t = \mu(v), \text{ $v$ is a subword of $u$} \right\}.
	\end{equation*}
	We claim that every conelike with respect to $\tau$ is contained in such a pair. Indeed, let $x = \nu(u)$, and let $w \in \nu^{-1}(\uparrow x)$. Let $\tilde{w}$ be a $\mu$-minor of $w$. Since $\tilde{w}$ is a subword of $v$ it is also a subword of $u$ and thus $\mu(w) = \mu(\tilde{w}) \in S$. Since $u$ was arbitrary, the result follows.
\end{proof}

\section{From Saturations to Conelikes}
\label{sec:EasyDirection}

In this section, we prove that all sets in the sets of subsets in Theorem \ref{thm:Pointlikes} are pointlikes and conelikes respectively. The characterization of the pointlikes for $\mathbf{J}_1$ follows immediately using Lemma \ref{lem:FixedAlphabet}. Because it is trivial, we show both directions directly.

\begin{proof}[Proof of Theorem \ref{thm:Pointlikes} \itref{aaa:PointlikesForRandLLevels}]
	The set of pointlikes contains the singletons and is closed under multiplication and subsets. Let $\relmorsmall{\tau}{A^*}{M}{N}{\mu}{\nu}$ where $A$ is the alphabet corresponding to the content morphism of $M$. By Lemma \ref{lem:FixedAlphabet}, we need only consider relational morphisms of this form. Furthermore, we can assume that $N$ has a content morphism $\beta: N \to J_A$. Indeed, we can if necessary consider $\relmorsmall{\tau'}{A^*}M{N \times J_A}\mu{\nu'}$ where $\nu'(a) = (\nu(u),\alp(u))$ for all $u \in A^*$. It is clear that if $S \subseteq M$ is pointlike with respect to $\tau'$, it is also pointlike with respect to $\tau$.

	Let $S \in \sat{\mathbf{J}_1}{M}$. Since $\alpha(s) = \alpha(t)$ for all $s,t \in S$, we have $\beta(x) = \beta(y)$ for all $x \in \tau(s)$, $y \in \tau(t)$. In other words, $x$ and $y$ are generated by the same elements in $A^*$. In $\mathbf{J}_1$, this implies that they are indeed the same, and thus $S$ is pointlike with respect to $\tau$. 
	For the other direction, we note the trivial inclusions $\pl{\mathbf{J}_1}{M} \subseteq \pl{\alpha}{M} = \sat{\mathbf{J}_1}{M}$.
\end{proof}

Next, we consider our final special case, that of $\mathbf{J}$. The idea here is that inside $\mathbf{J}$, any two idempotents with the same alphabet are the same element. Furthermore, any element which has an idempotent as factor is itself idempotent.

\begin{lemma}
	\label{lem:Jeasydirection}
	Let $M$ be a monoid with a content morphism $\alpha: M \to J_A$. Then $\sat{\mathbf{J}}{M} \subseteq \pl{\mathbf{J}}{M}$.
\end{lemma}

\begin{proof}
	We use induction over the construction of sets in $\sat{\mathbf{J}}M$. Similar to the proof of Lemma \ref{lem:inductionRandL}, all we need to show is that $XEY \in \pl{\mathbf{J}}{M}$ where $X$, $Y$ and $E$ has the properties in the definition of $\sat{\mathbf{J}}{M}$.
	
	Let $\relmorsmall\tau{A^*}M{N \in \mathbf{J}}\mu\nu$ be a relational morphism. As in the proof of Theorem \ref{thm:Pointlikes} \itref{aaa:PointlikesForRandLLevels}, we can assume that there is a content morphism $\beta: N \to J_A$. Let $R$ be the Ramsey number of $N$ and consider $u \in E$. Since $E$ is idempotent, there exists a factorisation $u = u_1 \dots u_{R-1}$ where $u_i \in E$. In particular, $\alpha(u_i) = \alpha(u)$ for each $i$. Choose $v_i \in \tau(u_i)$. By an argument similar to that in Lemma \ref{lem:longwordsinminors}, there exists a factor $e = v_i \dots v_j$ which is idempotent. We have 
	\begin{equation*}
		(v_1 \dots v_{i-1} e v_{j+1} \dots v_R) (v_1\dots v_{i-1} e v_{j+1} \dots v_R) = v_1 \dots v_{i-1} e v_{j+1} \dots v_R
	\end{equation*}
	by Lemma \ref{lem:daproperty}. Thus $e' = v_1 \dots v_{i-1} e v_{j+1} \dots v_R \in \tau(u)$ is also idempotent.

	Let $u' \in E$, and let $f \in \tau(u')$ be an idempotent obtained as above. Let $x \in X$, $y \in Y$ with $\hat{x} \in \mu^{-1}(x)$, $\hat{y} \in \mu^{-1}(y)$. Again by Lemma \ref{lem:daproperty}, we get $\nu(\hat{x})fe'f\nu(\hat{y}) = \nu(\hat{x})f\nu(\hat{y})$ and $e'\nu(\hat{x})f\nu(\hat{y})e' = e'$ showing that $\nu(\hat{x})f\nu(\hat{y}) \Jeq e'$ which implies $\nu(\hat{x})f\nu(\hat{y}) = e'$ by \greenJ-triviality. Since $x$, $u'$ and $y$ were arbitrary in their respective sets, we get $XEY \subseteq \tau^{-1}(e')$ showing that $XEY$ is pointlike. 
\end{proof}

We continue with the inductive cases. The varieties $\mathbf{R}_m$ and $\mathbf{L}_m$ are handled first since the subsequent cases depend on these results.

\begin{lemma}\label{lem:inductionRandL}
	Let $M$ be a monoid with a content morphism $\alpha: M \to J_A$, and let $m \geq 2$. Then $\satR{m}M \subseteq \plR{m}M$ and $\satL{m}M \subseteq \plL{m}M$.
\end{lemma}

\begin{proof}
	We will proceed by induction over $m$ and by structural induction over the construction of sets in $\satR{m}M$ and $\satL{m}M$. We only show the result for $\satR{m}M$; the result for $\satL{m}M$ follows by symmetry.
	Since pointlikes are closed under subsets and multiplication, all we need to show is that if $U \in \satR{m}M$ is idempotent and pointlike with respect to $\mathbf{R}_{m}$, $Z \in \satL{m-1}{M}$ is pointlike with respect to $\mathbf{L}_{m-1}$ and $\alpha(Z) \leq \alpha(U)$ then $UZ$ is pointlike with respect to $\mathbf{R}_{m}$
	
	Let $\relmorsmall\tau{A^*}M{N \in \mathbf{R}_m}\mu\nu$ be a relational morphism, and suppose that $\beta: N \to J_A$ is a content morphism. As by the previous proofs, this is all relational morphisms which we need to consider.
	
	By structural induction, we have $x \in N$ witnessing that $U$ is pointlike. Since $U$ is idempotent, have $\tau^{-1}(x^{\omega_N}) \supseteq (\tau^{-1}(x))^{\omega_N} \supseteq U^{\omega_N} = U$ and thus we can assume that the witness $x$ is idempotent.
	
	By induction over $m$, we have an equivalence class $Y$ over $\sim_{\mathbf{K}}$ such that $Z \subseteq \tau^{-1}(Y)$. Let $s \in U$, $t \in \tau^{-1}(y)$ where $y \in Y$. We have $\beta(y) = \alpha(t) \leq \alpha(s) = \beta(x)$. By Lemma \ref{lem:daproperty}, we have $xyx = x$ and since $y_i \sim_{\mathbf{K}} y$ for all $y_i \in Y$ it follows that $xy_i = xy$ for all such $y_i$. Hence $UZ$ is pointlike with $xy$ as a witness.
\end{proof}

For the other levels, we recall that the introduced closed sets build on $\mathbf{RL}_m$ factors. Our first step is to show that these are pointlike with respect to $\mathbf{R}_{m+1} \cap \mathbf{L}_{m+1}$ which in particular implies that they are pointlike with respect to $\Si_m$, $\Pivar_m$ and $\mathbf{R}_m \vee \mathbf{L}_m$.

\begin{lemma}\label{lem:RLSetsArePointlike}
	Let $M$ be a monoid with a content morphism $\alpha: M \to J_A$, and let $U = XEWFY$ be an $\mathbf{RL}_m$-factor of $M$. Then $U \in \plFO{m}{M}$ with an idempotent witness.
\end{lemma}

\begin{proof}
	Let $\relmorsmall\tau{A^*}M{N \in \mathbf{R}_{m+1} \cap \mathbf{L}_{m+1}}\mu\nu$ be a relational morphism and suppose $\beta: N \to J_A$ is a content morphism.
	Since $N \in \mathbf{L}_{m+1}$ and $E \in \satR{m}M$, there exists $D \subseteq N$ such that $D$ is a conjugacy class over $\sim_{\mathbf{D}}$ and $E \subseteq \tau^{-1}(D)$. We let
	\begin{equation*}
		X = \left\{ x = x_1 \cdots x_n \mid n \geq 1,  x_i \in D, x^2 = x \right\}.
	\end{equation*}
	Note that $X$ is also a conjugacy class over $\sim_{\mathbf{D}}$. Indeed, if $x,x_i \in X$, then $x^{n\omega_N} \sim_{\mathbf{D}} (x_1 \cdots x_n)^{\omega_N}$ by stability. Let $s \in \tau^{-1}(D)$. We have $s = s_1 \cdots s_n$ where $n$ is arbitrary and $s_i \in E$. Using a Ramsey argument similar to that of Lemma \ref{lem:longwordsinminors}, we get an idempotent factor $s' = s_j \cdots s_k$ of $s$. Since $s_i \in E$ and $E$ is idempotent, it follows that $s' \in E$. Thus there exist $x \in \tau(s') \cap D$. We get $x^{\omega_N} \in \tau(s')^{\omega_N} \subseteq \tau(s')$. Furthermore, let $x_i \in \tau(s_i)$. Since $\beta(x_i) = \beta(x^{\omega_N})$, it follows from Lemma \ref{lem:daproperty} that
	\begin{align*}
		x_1 \cdots x_{j-1} x^{\omega_N} x_{k+1} \cdots x_n x_1 \cdots x_{j-1} x^{\omega_N} x_{k+1} \cdots x_n = x_1 \cdots x_{j-1} x^{\omega_N} x_{k+1} \cdots x_n
	\end{align*}
	Thus $x_1 \cdots x_{j-1} x^{\omega_N} x_{k+1} \cdots x_n \in \tau(s)$ is idempotent. Since $s$ was arbitrary, we have $E \subseteq \tau^{-1}(X)$. We also get a conjugacy class $P$ over $\sim_{\mathbf{D}}$ such that $S \subseteq \tau^{-1}(P)$.

	Similarly, we get a conjugacy class $Y$ over $\sim_{\mathbf{K}}$ such that $Y$ contains only idempotents and such that $F \subseteq \tau^{-1}(Y)$, and we get a conjugacy class $Q$ over $\sim_{\mathbf{K}}$ such that $T \subseteq \tau^{-1}(Q)$.
	Finally, let $Z = \tau(W)$.
	Let $x,x' \in X$, $y,y' \in Y$, $p, p' \in P$, $s, s' \in Q$ and $z,z' \in Z$. Since $\beta(p), \beta(q), \beta(z) \leq \beta(x) = \beta(y)$ we get
  \begin{equation*}
	  p x z y q =  p x y x z y q = p x y q = p x y' q'  = p' x' y' q' = p' x'z'y' q'
  \end{equation*}
  where we used Lemma \ref{lem:daproperty} and the fact that $x \sim_{\mathbf{D}} x'$, $y \sim_{\mathbf{K}} y'$, $p \sim_{\mathbf{D}} p'$ and $q \sim_{\mathbf{K}} q'$
  Since these choices were arbitrary, we get that $PXZYQ$ contains a single element, showing that $SEWFT$ is pointlike with the unique element of $PXZYQ$ as witness. Furthermore, again by Lemma \ref{lem:daproperty}, $pxzyqpxzyq = pxzyq$, so the witness is idempotent.
\end{proof}

We are now ready to prove inclusion of the introduced closed sets in the pointlikes/conelikes. We give the Lemma only for $\Si_m$; analogous lemmas for the other levels follows analogously.

\begin{lemma}\label{lem:satispointlikecone}
	Let $M$ be a monoid with a content morphism $\alpha: M \to J_A$ and let $m \geq 2$. Then $\psatS{m}{M} \subseteq \coneS{m}{M}$.
\end{lemma}

\begin{proof}
	Let $\relmorsmall\tau{A^*}M{N \in \Si_m}\mu\nu$ be a relational morphism and let $\beta: N \to J_A$ be a content morphism. We need to show that 
	\begin{equation*}
		(u_1,U_1) (v_1,V_1) (u_2,U_2) \dots (v_{n-1},V_{n-1}) (u_n,U_n) \in \cone{\tau}M
	\end{equation*}
	where $U_i$ is an $\mathbf{RL}_m$-factor, $u_i \in U_i$, $(v_i,V_i) \in \psatS{m-1}M$ and $\alpha(v_i) \leq \alpha(U_i), \alpha(U_{i+1})$ for all $v_i \in V_i$. We proceed by induction on $m$.

	Since $N \in \Si_m$, it follows in particular that $N$ with the order removed is in $\mathbf{R}_{m+1} \cap \mathbf{L}_{m + 1}$. Thus Lemma \ref{lem:RLSetsArePointlike} shows that for all $1 \leq i \leq n$, there exists $e_i$ such that $U_i \subseteq \tau^{-1}(e_i)$. By induction, we have that for all $1 \leq i \leq n-1$ there exists an element $z_i \in \tau(v_i)$ and a set $Z_i \subset N$ such that $V_i \subseteq \tau^{-1}(Z_i)$ and $z_{i} \preceq_{\mathbf{KD}} z_i'$ for all $z_i' \in Z_i$.

	Let $x' \in U_1 V_1 U_2 \dots V_{n-1}U_n$, and let $z_i' \in Z_i$ such that $x'\in \tau^{-1}(e_1 z_1' e_2 \dots e_{n-1}z_{n-1}'e_{n})$. 
	Since $\beta(z_i') \leq \beta(e_i),\beta(e_{i+1})$ it follows from Lemma \ref{lem:daproperty} that $e_i z_i' e_i = e_i$ and $e_{i+1} z_i' e_{i+1} = e_{i+1}$. 
	It follows from the definition of $\preceq_{\mathbf{KD}}$ that
	\begin{align*}
		e_1 z_1 e_2 z_2 \dots z_{n-1} e_n & \leq e_1 z'_1 e_2 z_2 \dots z_{n-1} e_n
		\\ & \leq e_1 z'_1 e_2 z'_2 \dots z_{n-1} e_n
		\\ & \;\,\vdots
		\\ & \leq e_1 z'_1 e_2 z'_2 \dots z'_{n-1} e_n.
	\end{align*}
	Since $x'$ was arbitrary, this shows that $U_1V_1 \dots U_{n-1}V_{n-1}U_{n}$ is conelike with the witness $e_1z_1e_2z_2 \dots z_{n-1}e_n$.
\end{proof}

\section{On the Structure of Comparison Definable Sets}
\label{sec:OnStructure}

In this section, we explore some structural properties of particular sets defined by ranker comparisons. We will consider $\RR_{1,n} = \XY_{2,n} \cap \YX_{2,n} = \left\{ (r,s) \in R_{1} \times R_{1} \mid |r|,|s| \leq n \right\}$, i.e.~all comparisons where the rankers has alternation depth $1$ and length $n$.
The properties of these sets are used in subsequent sections to show that some $\XY_{m,n}$ or $\YX_{m,n}$-sets are in the relevant closed set. We also consider a special case which does not contain $\RR_{1,n}$, namely $\XY_{1,n} \cup \YX_{1,n}$.

We will show how these languages consist of certain \emph{long} factors, intuitively factors which are long enough so that the relevant rankers can not see the whole factor. We use the following notation:
If $u = a_1 \dots a_n$ and $1 \leq i \leq j \leq n$, then $u[i] = a_i$, and $u[i,j) = a_{i} \dots a_{j-1}$. Note in particular that $u[i,i)$ is the empty factor.

\begin{definition}
	Let $u \in A^*$. We say that a factor $u[i,j)$ is \emph{$n$-long} if every word $v \in \alp(u[i,j))^*$ such that $|v| \leq n$ is a subword of $u[i,j)$. Note in particular that this definition vacuously implies that empty factors $u[i,i)$ are $n$-long. Given a word $u$, we call a factor $u[i,j)$ \emph{maximal $n$-long} if it is $n$-long and for every $n$-long $u[i',j')$ such that $i' \leq i \leq j \leq j'$ we have $i' = i$ and $j' = j$. In other words, an $n$-long factor is maximal if it is not properly contained in any other $n$-long factor.
\end{definition}

\begin{definition}
	Let $u \in A^*$. We define 
	\begin{equation*}
		\mathsf{R}_n(u) = \left\{ 0 \right\} \cup \left\{ i \in \mathbb{N} \mid  \text{$u[j,i)$ is a maximal $n$-long factor for some $j$} \right\}
	\end{equation*}
	Note that since the factor $u[j,i)$ does not contain the position $i$, the set $\mathsf{R}_u(u)$ consists of the positions just after some maximal $n$-long factor. Symmetrically, we define
	\begin{equation*}
		\mathsf{L}_n(u) = \left\{ |u| + 1 \right\} \cup \left\{ i \in \mathbb{N} \mid  \text{$u[i+1,j)$ is a maximal $n$-long factor for some $j$} \right\}.
	\end{equation*}
\end{definition}
	Note that the maximality of $u[j,i)$ implies that $u[i] \notin \alp(u[j,i))$ in the definition of $\mathsf{R}_n(u)$ and symmetrically $u[i] \notin \alp(u[i+1,j))$ in the definition of $\mathsf{L}_n(u)$.
	For $n = 1$, any factor is long, and thus $\mathsf{R}_n(u) = \mathsf{L}_n(u) = \left\{ 0, |u| + 1 \right\}$. In this case, the following results are trivial, and in order to avoid dealing explicitly with it, we always assume $n \geq 2$.

	We want to show that in between each pair of positions in $\mathsf{R}_n(u)$, there is a position in $\mathsf{L}_n(u)$ (possibly coinciding with one of the positions in $\mathsf{R}_n(u)$) and vice versa. First, we prove the following Lemma.

\begin{lemma}
	Suppose $\left\{ i, i + 1, \dots ,j - 1, j \right\} \cap \mathsf{R}_n(u) > 1$, then $u[i+1,j+1)$ is not $n$-long.
	\label{lem:onlyoneaineachlongfactor}
\end{lemma}

\begin{proof}
	For every $n$-long factor, we can consider the maximal $n$-long factor containing it. Thus, we lose no generality in proving the statement only for maximal $n$-long factors. Assume $k,\ell \in \mathsf{R}_n(u)$ with $i \leq k < \ell \leq j$ and assume $u[i+1,j+1)$ is maximal $n$-long. We have a maximal $n$-long $u[\ell',\ell]$. Since $\ell < j + 1$, we must have $\ell' < i + 1$; otherwise $u[\ell',\ell]$ would be properly contained in $u[i+1,j+1)$ and thus could not be maximal $n$-long. In particular, this leads to a contradiction if $i = 0$, since $\ell' < 1$ is impossible.
	
	If $i \geq 1$, then there exists a maximal $n$-long $u[k',k]$. In particular, $k' < \ell'$. Since $u[k] \in \alp(u[\ell',\ell]) \setminus \alp(u[k',k])$ we must have $\alp(u[k+1,\ell]) = \alp(u[\ell',\ell])$, since the subword $u[k]q$ with $q \in \alp(u[\ell',\ell])^{n-1}$ must exist in $u[\ell',\ell]$. Thus $\alp(u[\ell',\ell]) \subseteq \alp(u[i+1,j+1))$. Since these factors intersect and are maximally $n$-long, they must coincide which contradicts $\ell \leq j$.
\end{proof}

The above Lemma says that even if maximal $n$-long factors are not necessarily disjoint (note e.g.\ $(ab)^na(ac)^{n}$), no $n$-long factor can cover more than one $\mathsf{R}_n$-marker.\footnote{In fact, it says something slightly stronger. Even if we include the position just before an $n$-long factor, this set of positions can not cover more than one $\mathsf{R}_n$-marker.} This gives us the following Lemma.

\begin{lemma}\label{lem:intersectionsoflongsubwords}
	Let $i,j \in \mathsf{R}_n(u)$ with $i < j$. Then there exists $k \in \mathsf{L}_n(u)$ such that $i < k \leq j$. Symmetrically, if $i,j \in \mathsf{L}_n(u)$ with $i < j$, then there exists $k \in \mathsf{R}_n(u)$ with $i \leq k < j$.
\end{lemma}

\begin{proof}
	Let $a[k+1,\ell]$ be a maximal $n$-long factor containing $a[j+1,j+1)$. By definition, $k \leq j$ and by Lemma \ref{lem:onlyoneaineachlongfactor}, we must have $i < k$, giving the desired result. The other direction is symmetrical.
\end{proof}

We will now turn to the factorizations of $\RR_{1,n}$-sets. We start with factorizing a single word into the desired form. 

\begin{lemma}\label{lem:structurefactorizationofsingleword}
	Given $u \in A^*$ and $n \geq 1$, there is a factorization
	\begin{equation} \label{eqn:structurefactorizationofsingleword}
		u = u_1 b_1 v_1 a_1 \dots u_{k-1} b_{k-1} v_{k-1} a_{k-1} u_k,
	\end{equation}
	where $v_i$ can be empty, $a_i$ and $b_i$ can coincide, such that for all $i$
	\begin{enumerate}[(i)]
		\item $u_i$ is $n$-long, \label{aaa:structurefactorizationofsingleword}
		\item $\alp(v_i) \subsetneq \alp(u_ib_i) \cap \alp(a_iu_{i+1})$,\label{bbb:structurefactorizationofsingleword}
		\item $a_i$ is reachable by a ranker $\mathsf{X}_{p_i}$ and $b_i$ by a ranker $\mathsf{Y}_{q_i}$ where $|p_i|,|q_i| \leq (n+1)|A\setminus \alp(v_i)|$,\label{ccc:structurefactorizationofsingleword}
		\item either the markers $a_i$ and $b_i$ coincide or $a_i \in \alp(u_{i+1})$, $b_i \in \alp(u_{i})$.\label{ddd:structurefactorizationofsingleword}
	\end{enumerate}
\end{lemma}

\begin{proof}
	By Lemma \ref{lem:intersectionsoflongsubwords}, we can choose $a_i$ to mark the positions from $\mathsf{R}_{n+2}(u)$ and $b_i$ to mark the positions from $\mathsf{L}_{n+2}(u)$, and get a factorization with the markers $a_i$ and $b_i$ interlaced, or possibly coinciding, as in (\ref{eqn:structurefactorizationofsingleword}).
	We want to show that conditions (\ref{aaa:structurefactorizationofsingleword} - \ref{ddd:structurefactorizationofsingleword}) hold for this factorization. 

	Let $i$ be fixed. Since $u_i$ contains no position in $\mathsf{R}_{n+2}(u)$ or $\mathsf{L}_{n+2}(u)$, it is contained in some maximal $(n+2)$-long factor, say $w_i$. By the definition of $\mathsf{R}_{n+2}(u)$ and $\mathsf{L}_{n+2}(u)$, the marker $b_{i-1}$ must be directly to the left of $w_i$, and the marker $a_i$ need to be directly to the right of $w_i$. In other words, we have one of the following cases:
	\begin{enumerate}[(a)]
		\item $w_i = u_i$, if $b_{i-1}$ coincides with $a_{i-1}$ or $i = 1$ and $b_{i}$ coincides with $a_i$ or $i = k$,
		\item $w_i = u_ib_iv_i$, if $b_{i-1}$ and $a_{i-1}$ coincide or $i = 1$ while $a_i$ and $b_i$ do not coincide,
		\item $w_i = v_{i-1}a_{i-1}u_i$, if $b_{i}$ and $a_{i}$ coincide or $i = k$ while $a_{i+1}$ and $b_{i+1}$ does not coincide,\label{ccc:factorproof}
		\item $w_i = v_{i-1}a_{i-1}u_ib_{i}v_i$ otherwise.\label{ddd:factorproof}
	\end{enumerate}

	We will show properties \itref{aaa:structurefactorizationofsingleword},\itref{bbb:structurefactorizationofsingleword} and \itref{ddd:structurefactorizationofsingleword} for case \itref{ddd:factorproof}. The others are handled similarly. Since $1 < i$, we have that $w_{i-1}$ exist. Since it is maximal $(n+2)$-long, we must have $a_{i-1} \notin \alp(w_{i-1}) \supseteq \alp(v_{i-1})$. In particular, it follows from the fact that $w_i$ is $(n + 2)$-long that $u_ib_iv_i$ is $(n + 1)$-long with $\alp(w_i) = \alp(u_ib_iv_i)$. Applying the same argument from the right, shows that $u_i$ is $n$-long with $\alp(u_i) = \alp(w_i)$. This shows \itref{aaa:structurefactorizationofsingleword} and \itref{ddd:structurefactorizationofsingleword}. Furthermore, the fact that $b_i \notin \alp(w_{i+1}) \supseteq \alp(v_i)$ shows condition \itref{bbb:structurefactorizationofsingleword}.

	Thus, all that is left is to show condition \itref{ccc:structurefactorizationofsingleword}. By symmetry, we only show that $a_i$ is reachable by a ranker $\mathsf{X}_{p_i}$ with $|p_i| \leq (n+1)|A\setminus \alp(v_i)|$. We will use induction on the size of the alphabet.

	If $A = \left\{ a \right\}$ then either $u = u_1$ or $u = a^{m}$ where $m \leq n + 1$. In the former case, there are no markers $a_i$, and thus nothing to show.
	In the latter case, we note that every position is both an $a_i$ and a $b_i$ marker. Thus, every $v_i$ is empty. If $a_i$ is the $\ell$\th position of $u$, then $\mathsf{X}_{a^{\ell}}$ is a ranker of the desired length.
	
	Let $|A| \geq 2$ and let $c$ mark the rightmost position which is not to the right of $a_i$ and which is reachable by a ranker $\mathsf{X}_{p}$ where $|p| \leq n + 1$. Let $a_{j}$ be the leftmost marker occurring after or at $c$, let $s$ be the factor of $u$ up to the marker $c$ and let $t$ be the factor between $c$ and $a_{j}$. We define $u' = t a_{j} \dots u_{i}b_{i}v_{i}$.
	
	Note that $a_i \notin \alp(w_i)$, $b_i \notin \alp(w_{i+1})$. If these markers do not coincide, then $a_i \in \alp(w_{i+1})$ and $b_i \in \alp(w_i)$, showing that $\alp(v_i) \leq |A| - 2$. Furthermore, if $a_i$ and $b_i$ do coincide, then $v_i$ is empty. Since $|A| \geq 2$, we have $|A \setminus \alp(v_i)| \geq 2$ in both cases.
	In particular, if $a_i \notin \alp(u')$, then the ranker $\mathsf{X}_{pa_i}$ has the desired properties.

	Next, let us assume that $a_i \in \alp(u')$. In order to use induction, we need to show that $\alp(u') \subsetneq \alp(u)$. 
	Since $a_i \in \alp(u')$ and $a_i \notin \alp(w_i)$, the factor $u'$ must begin before the factor $w_i$. In particular, this means that $scu'$ is not $(n+2)$-long (since $w_i$ is maximally $n+2$-long). Thus, there exists $r \in A^*$, $d \in A$ with $|r| \leq n + 1$ such that $\mathsf{X}_{r}(u) \leq |scu'|$ and either $|scu'| < \mathsf{X}_{rd}(u)$ or $\mathsf{X}_{rd}$ is undefined on $u$. By the definition of $p$, we have, $\mathsf{X}_r(u) \leq \mathsf{X}_{p}(u) = |sc|$. Thus, $d \notin \alp(u')$.
	
	Since $w_i$ is contained in $u$, it must be maximal $(n+2)$-long in $u'$. Since $a_i \notin \alp(w_i)$, the word $w_i$ is maximally $(n+2)$-long also in $u'a_i$. It follows that $a_i$ marks a position in $\mathsf{R}_{n+2}(u'a_i)$. By induction, there exists $\mathsf{X}_{p'}$ where $|p'| \leq (n+1)|\alp(u')\setminus \alp(v_i)|$ such that $\mathsf{X}_{p'}(u'a_i) = |u'a_i|$. Thus the ranker $\mathsf{X}_{pp'}$ has the desired properties. 
\end{proof}

\newcommand{\SPrank}[2]{\RR_{#1,#2}}
\newcommand{\SPrankwn}{\SPrank{m}{(n+1)|A| + n}}

Thus, we can obtain a factorization with some desired properties for each word $u$. The following proposition shows that these factorizations  can be combined to give a factorization of a $\SPrankwn$-set.

\begin{lemma}\label{lem:structurefactorization}
	Let $n \in \mathbb{N}$ and let $U \in A^{*}$ be a $\SPrankwn$-set. Then it is possible to find a factorization
	\begin{equation}\label{eqn:structurefactorization}
		U = U_1 b_1 V_1 a_1 \dots U_{k-1} b_{k-1} V_{k-1} a_{k-1} U_k,
	\end{equation}
	where
	\begin{enumerate}[(i)]
		\item Each $U_i$ is $n$-long\label{aaa:structurefactorization}
		\item $\alp(V_i) \subsetneq \alp(U_ib_i) \cap \alp(a_{i}U_{i+1})$,\label{bbb:structurefactorization}
		\item For each $u \in U$, the position marked by $a_i$ is reachable by $\mathsf{X}_{p_i}$ and the position marked by $b_i$ is reachable by $\mathsf{Y}_{q_i}$ where $|p_i|,|q_i| \leq (n+1)|A \setminus \alp(V_i)|$, \label{ccc:structurefactorization}
		\item For each $i$, either the markers $a_i$ and $b_i$ coincide, or $a_i \in \alp(U_{i+1})$, $b_i \in \alp(U_{i})$.\label{ddd:structurefactorization}
	\end{enumerate}
\end{lemma}

\begin{proof}
	Suppose $u \in U$ and factor it as Lemma \ref{lem:structurefactorizationofsingleword}. We note that $\mathsf{Y}_{q_i}(u) \leq \mathsf{X}_{p_i}(u) < \mathsf{Y}_{q_{i+1}}(u)$ which implies that the same must be true on $u'$. In particular, this gives the factorization
	\begin{equation*}
		u' =  u_1' b_1 v_1' a_1 \dots u_{k-1}' b_{k-1} v_{k-1}' a_{k-1} u_k'
	\end{equation*}
	We combine these factorizations for all $u' \in U$ to get a factorization of the form in (\ref{eqn:structurefactorization}). This factorization satisfies \itref{ccc:structurefactorization} by definition.

	Let $1 \leq i \leq k-1$ and let $r \in \alp(u_i)^*$ with $|r| \leq n$. Since every $u_i$ in the factorization of $u$ is $n$-long, we have that $\mathsf{X}_{p_{i-1}r}(u) < \mathsf{Y}_{q_i}(u)$ (if $i = 1$ we set $p_{i-1} = \varepsilon$). Thus the same is true for $u'$. Furthermore, if $a \notin \alp(u_i)$, then either $\mathsf{X}_{p_{i-1}a}$ is not defined on $u$ or $\mathsf{Y}_{q_i}(u) \leq \mathsf{X}_{p_{i-1}a}(u)$. Again, the same is true on $u'$. It follows that $\alp(u_i) = \alp(u_i')$ and $u_i'$ is $n$-long. For $i = k$, we consider the rankers $\mathsf{X}_{p_k}$ and $\mathsf{Y}_{\overline{r}}$ and get the same result. Hence the factorization satisfies \itref{aaa:structurefactorization}.

	Next, let $1 \leq i < k-1$. We have that $a_i$ and $b_i$ coincide on $u$ if and only if $\mathsf{X}_{p_i}(u) \leq \mathsf{Y}_{q_i}(u)$ and $\mathsf{Y}_{q_i}(u) \leq \mathsf{X}_{p_i}(u)$. Thus, this is true on $u$ if and only if it is true on $u'$.
	If $a_i$ and $b_i$ does not coincide, we get $a_i \in \alp(u_{i+1}) = \alp(U_{i+1})$ and $b_i \in \alp(u_i) = \alp(U_i)$ by Lemma \ref{lem:structurefactorizationofsingleword}. Thus the factorization satisfy \itref{ddd:structurefactorization}.
	
	If $a_i$ and $b_i$ does not coincide and $c \in \alp(v_i') \subseteq \alp(u_i'b_iv_i')$, then $\mathsf{X}_{p_{i}c}(u') < \mathsf{X}_{p_{i+1}}(u')$ which is also satisfied on $u$. Thus $c \in \alp(u_ib_iv_i)$, and by property \itref{bbb:structurefactorizationofsingleword} of Lemma \ref{lem:structurefactorizationofsingleword}, we get $c \in \alp(u_i) = \alp(U_i)$. Symmetrically, we get that if $c \in \alp(v'_i) \subseteq \alp(v'_ia_iu'_{i+1})$, then $c \in \alp(U_{i+1})$. This shows \itref{bbb:structurefactorization}.
\end{proof}

We also provide the following factorization which deals with the $\XY_{1,n} \cup \YX_{1,n}$-case.

\begin{lemma}\label{lem:structureofsubwordsets}
	Given an alphabet $A$ and integer $n$, and an $\XY_{1,(n+1)|A|+n} \cup \YX_{1,(n+1)|A| + n}$-set $U$, it is possible to find a factorization
	\begin{equation*}
		U = U_1 \lambda_1 U_2 \dots \lambda_{k-1} U_{k} \lambda_k  U_{k+1}
	\end{equation*}
	where $\lambda_i \in A \cup \left\{ \varepsilon \right\}$  i.e.\ the $\lambda_i$ are either letters or empty, and where each $U_i$ is $n$-long.
\end{lemma}

\begin{proof}
	Choose $u \in U$ and factor it according to Lemma \ref{lem:structurefactorizationofsingleword}. Let $\mathsf{X}_{p_i}$ and $\mathsf{Y}_{q_i}$ be the rankers from the Lemma, and let $u' \in U$. We factor
	\begin{equation*}
		u' = u'_1 b_1 v'_1 a_1 \dots u'_{k-1} b_{k-1} v'_{k-1} a_{k-1} u'_k
	\end{equation*}
	where $a_i$ marks the position of $\mathsf{X}_{p_i}$ and $b_i$ marks the position of $\mathsf{Y}_{q_i}$. For each $i$ we have words $p_{i}$ and $\overline{q}_{i+1}$ such that $a_i$ marks the last letter of the first occurrence of $p_i$ and $b_{i+1}$ marks the first letter of the last occurrence of $\overline{q}_i$. We set $p_0 = \varepsilon$ and $\overline{q}_k = \varepsilon$.

	For $1 \leq i < k-1$, suppose $p_i = p_i'a_i$ and $q_i = b_iq_i'$. We have that $a_i$ and $b_i$ coincide if and only if they are the same letter, $p_i'a_iq_i'$ is a subword of $u$ and $p_i'b_ia_iq_i'$ is not a subword of $u$. In particular, $a_i$ and $b_i$ coincide on $u'$ if and only if they coincide on $u$.

	Let $1 \leq i \leq k$ and let $r \in \alp(u_i)^*$ with $|r| \leq n$. Since every $u_i$ in the factorisation of $u$ is $n$-long, we have that $p_{i-1}r\overline{q}_{i}$ is a subword of $u$. Thus the same is true for $u'$. Furthermore, if $a \notin \alp(u_i)$, then $p_{i-1}a\overline{q}_i$ is not a subword of $u$ and hence neither of $u'$. It follows that $\alp(u_i) = \alp(u_i')$ and $u_i'$ is $n$-long.

	Suppose that $a_i$ and $b_i$ does not coincide. We factor $a_i v_{i}' b_i = s_it_i$ where $s_i$ is the longest factor such that $\alp(s_i) \subseteq \alp(u_i)$. Let $c$ be the first letter of $t_i$, and let $d \in \alp(u_i) \setminus \alp(u_{i+1})$.
	By the alphabetic requirements, we have that $p_{i-1}cd\overline{q}_i$ can not be a subword of $u$, and thus neither of $u'$. Hence $\alp(t_i) \subseteq \alp(u_{i+1})$.

	If $a_i$ and $b_i$ do coincide, we put $\lambda_i = a_i$, and $s_i = t_i = \varepsilon$. This way, we get a factor $t_{i-1}u_i's_i$ from each $u' \in U$ and each $i$. Setting $U_i$ to be the union of these factors gives us a factorisation of the desired form.
\end{proof}

\section{From Conelikes to Saturations}
\label{sec:DifficultDirection}

We now have everything we need to show that we can find monoids in the relevant varieties whose pointlikes/conelikes coincide with the introduced closed sets. Since the pointlikes/conelikes with respect to the variety is contained in the pointlikes/conelikes with respect to any particular monoid in the variety, this gives our desired result.

We start with the variety $\mathbf{J}$, where the result is obtained by two lemmas. First, we show that anything $(R-1)$-long is in $\sat{\mathbf{J}}M$. The result then follows trivially using the factorization in Lemma \ref{lem:structureofsubwordsets}.

\begin{lemma}\label{lem:FOpltosatHelpingLemma}
	Let $\mu: A^* \to M$ be a homomorphism, and let $R$ be the Ramsey number of $M$. If $U$ is $(R-1)$-long, then $\mu(U) \in \sat{\mathbf{J}}{M}$.
\end{lemma}

\begin{proof}
	Let $u \in U$. Since $u$ is $(R-1)$-long, we can factor $u = u_1 \dots u_{R-1}$ where $\alp(u_i) = \alp(U)$. By an argument similar to that of Lemma \ref{lem:longwordsinminors}, we find a factor $u_i \dots u_j$ which is idempotent. Setting $e = u_i \dots u_j$, we get a factorisation $u = xey$. Combining these factorisations for all $u \in U$, we get $U = XEY$ satisfying the alphabetic properties.

	Let $\omega$ be the idempotent power of $2^M$. Since every element in $E$ is idempotent, we have $XEY \subseteq XE^{\omega}Y$, where $E^{\omega}$ is clearly idempotent.
\end{proof}

\begin{lemma}\label{lem:FOspecialcase}
	Let $M$ be a monoid and let $R$ be the Ramsey number of $M$. Let $\relmorsmall{\tau}{A^*}{M}{N^{\XY \cup \YX}_{1,n} \in \mathbf{J}}{\mu}{\nu}$ with $n = |A|R+R-1$. Then $\pl{\tau}{M} \subseteq \sat{\mathbf{J}}{M}$.
\end{lemma}

 \begin{proof}
	 Let $s \in N^{\XY \cup \YX}_{1,n}$. We want to show that $\mu(\nu^{-1}(s)) \in \sat{\mathbf{J}}{M}$. To this end, we factor 
	 \begin{equation*}
		 \nu^{-1}(s) = U_1\lambda_1 \dots U_{k-1}\lambda_{k-1}U_{k}
	 \end{equation*}
	 according to Lemma \ref{lem:structureofsubwordsets} for $R-1$. We recall that each $a_i$ is either a single letter or empty. Thus, by closure under multiplication, it is enough to show that each $U_i \in \sat{\mathbf{J}}{M}$. Since each $U_i$ is $(R-1)$-long, this follows from Lemma \ref{lem:FOpltosatHelpingLemma}.
\end{proof}

For the varieties $\mathbf{R}_m$, $\mathbf{L}_m$ and for the $\mathbf{RL}_m$-sets, the idea is to choose the depth of the rankers long enough so that we can pump in the monoid $2^M$. It turns out that instead of asking for idempotents in $2^M$, a weaker condition is sufficient: being \emph{subidempotent}.

\begin{definition}
	Let $M$ be a monoid. A set $U \subseteq M$ is \emph{subidempotent} if $U \subseteq U^2$.
\end{definition}

We note that every subidempotent is, by definition, a subset of $U^{\theta}$ where $\theta = \omega_{2^M}$ is the idempotent power of $2^M$. Since pointlikes and conelikes are closed under subsets, this means that any mention of idempotent in Definition \ref{def:RLSaturations} and \ref{def:RLmFactors} can be substituted with subidempotent without changing the result. Furthermore, subidempotents are easily obtained by long enough factorizations.

\begin{lemma}\label{lem:ramseysapplied}
	Let $M$ be a monoid, and let $R$ be the Ramsey number of the set $2^M$. Let $U \subseteq A^*$ have the factorization $U = U_1 \dots U_n$, with $n \geq R-1$. Then there exists $1 \leq i,j \leq n$ such that $\mu(U_i U_{i+1} \dots U_{j-1} U_j)$ is subidempotent.
\end{lemma}

\begin{proof}
	The empty set is subidempotent, and thus the statement is true for the degenerate case $U = \emptyset$. We assume $U \neq \emptyset$ which implies $U_i \neq \emptyset$ for all $i$.

	Let $\mathcal{G} = (V,E)$ be the complete graph with vertices $V = \left\{ 1,\dots,n+1 \right\}$. The map $\mu$ induces an edge-coloring $c$ of size $2^{|M|}$ on $\mathcal{G}$ by $c(\left\{ i,j \right\}) = \mu(U_i \dots U_{j-1})$. By Theorem \ref{thm:ramseys}, there exists a monochrome triangle. In other words, there exists $i$, $j$, $k$ with the property that $\mu(U_i \dots U_{j-1}) = \mu(U_j \dots U_{j-1}) = \mu(U_i \dots U_{k-1})$. Let us call this set $S$. Since $S = \mu(U_i \dots U_{k-1}) \subseteq \mu(U_i \dots U_{j-1})\mu(U_j \dots U_{k-1}) = S^2$ it follows that $S$ is subidempotent.
\end{proof}

We now prove the desired result for $\mathbf{R}_m$ and $\mathbf{L}_m$. By obtaining a long enough factorization, we find an idempotent using the above Lemma. Everything before the idempotent is handled using Lemma \ref{lem:subwordsubrankerXX} and induction on the alphabet, whereas everything after is handled using the same lemma and induction on $m$.

\begin{lemma}\label{lem:inductionRandLother}
	Let $M$ be a monoid with a content morphism $\alpha: M \to J_A$. Let  $m \geq 1$, and let $\relmorsmall{\tau}{A^*}{M}{N^{\XX}_{m,n}}{\mu}{\nu}$ be the natural relational morphism where $n = (m+|A|)(R-1)$ depends on the Ramsey number $R$ of $2^M$. Then $\pl{\tau}{M} \subseteq \satR{m+1}{M}$. Dually, if $\relmorsmall{\tau'}{A^*}{M}{N^{\YY}_{m,n}}{}{}$, then $\pl{\tau'}{M} \subseteq \satL{m+1}{M}$. 
\end{lemma}

\begin{proof}
	By symmetry, we need only show the result for $N^{\XX}_{m,n}$. We show that if $U$ is an $\XX_{m,n}$-set, then $\mu(U) \in \satR{m-1}{M}$. In particular, this shows that $\mu(\nu^{-1}(s)) \in \satR{m+1}{M}$ for all $s \in N^{\XX}_{m,n}$, which is the desired result.
	We use induction over $(m,|A|)$ ordered alphabetically. 
	First, we show that we can factor $U$ into one of the following forms
\begin{align*}
	U & = U_1a_1U_2a_2 \cdots U_{k} a_k U_{k+1}
	\\ U & = U_1a_1U_2a_2 \cdots U_{(R-1)} a_{(R-1)} V
\end{align*}
subject to the conditions
\begin{enumerate}[(i)]
	\item $k < (R-1)$
	\item $\alp(U_i) \subsetneq \alp(U_ia_i) = A$ for all $i$ (if $U$ does not contain all letters in $A$, then $k = 0$).\label{bbb:RandLproof}
\end{enumerate}
Indeed, suppose we already factored $U = U_1 a_1 \cdots U_i a_i W$ such that $i < (R-1)$ and condition \itref{bbb:RandLproof} holds. In particular, $a_i$ is reachable by the ranker $\mathsf{X}_{a_1 \dots a_i}$ on all $v \in U$. If $i = R-1$, we choose $V = W$ and if $\alp(W) \subsetneq A$, we choose $U_{k+1} = W$. In both these cases, we are done. Thus, we suppose that $\alp(W) = A$ and that $i < (R-1)$. 

For some $u \in U$, choose the letter $a_{i+1}$ such that $\mathsf{X}_{a_1 \ldots a_{i} b}(u) \leq \mathsf{X}_{a_1 \ldots a_{i} a_{i+1}}(u)$ for all $b \in A$. It is clear that this choice is unique and since $U$ is an $\XX_{m,n}$-set, it is independent of the choice of $u$. We factor $U = Ua_1 \cdots U_ia_iU_{i+1}a_{i+1}W'$ where $a_{i+1}$ marks the first occurrence of $a_{i+1}$ after the position marked by $a_i$. By choice of $a_{i+1}$, we have $\alp(U_{i+1}) = A \setminus \left\{ a_{i+1} \right\}$ and so this factorization satisfies \itref{bbb:RandLproof} and is thus a factorization of the desired form.

Since $a_i$ is reachable by the ranker $\mathsf{X}_{a_1 \dots a_i}$, which in particular has depth at most $R-1$, it follows from Lemma \ref{lem:subwordsubrankerXX} that $U_i$ is an $\XX_{m,n'}$-set where $n' = (m + A - 1)(R-1) = (m + |\alp(U_i)|)(R-1)$. Let $M_i$ be the submonoid of $M$ generated by $\alp(U_i)$. Since the Ramsey number of $2^{M_i}$ is at most $R$, it follows from induction on the size of the alphabet that $\mu(U_i) \in \satR{m+1}{M_i} \subseteq \satR{m+1}{M}$.  Since closed sets are closed under multiplication, we get $\mu(U_ia_i \cdots U_j a_j) \in \satR{m+1}{M}$ for every factor $U_i a_i \dots U_j a_j$ of $U$. In particular, this gives $\mu(U) \in \satR{m+1}{M}$ when the factorization is of the first form.

For the second form of the factorization, we know by Lemma \ref{lem:ramseysapplied} that there exists $U_ia_i \cdots U_j a_j$ such that $\mu(U_i a_i \cdots U_j a_j)$ is subidempotent. By the argument above, it follows that we have $\mu(U_1 a_1 \dots U_{i-1} a_{i-1}),\mu(U_{i} a_{i} \dots U_{j} a_j) \in \satR{m+1}{M}$. Let $E = \mu(U_i a_i \cdots U_j a_j)^{\theta}$ where $\theta = \omega_{2^{M}}$ is the idempotent power of $2^M$. By subidempotency, we have $U_i a_i \cdots U_j a_j \subseteq E$.

Suppose $m \geq 2$. Since $a_j$ is reachable by a ranker $\mathsf{X}_{p_i}$ of length at most $R-1$, it follows by Lemma \ref{lem:subwordsubrankerXX} that $U_{j+1} a_{j+1} \dots U_{(R-1)}a_{(R-1)} V$ is a $\XX_{m,n''}$-set where $n'' = (m-1 + |A|)(R-1)$. This in particular implies that it is a $\YY_{m-1,n''}$-set. If $m = 1$, we can use definedness of rankers $\mathsf{X}_{a_1 \ldots a_i b}$ for $b \in A$ to conclude that $U_{j+1}a_{j+1} \cdots U_{(R-1)}a_{(R-1)}V$ has a well defined alphabet. Thus in both cases $\mu(U_{j+1}a_{j+1} \dots U_Ra_RV) \in \satL{m}{M}$ (in the former case by induction, and in the latter by definition). We get
\begin{align*}
	\mu(U) & \subseteq  \mu(U_{1} a_{1} \dots U_{i} a_i)\mu(U_{i} a_{i} \dots U_{j} a_j)\mu(U_{j+1} a_{j+1} \dots U_{(R-1)}a_{(R-1)}V) \\ & \subseteq \mu(U_{1} a_{1} \dots U_{i} a_i)E\mu(U_{j+1} a_{j+1} \dots U_{(R-1)}a_{(R-1)}V) \in \satR{m}M
\end{align*}
which is the desired result.
\end{proof}

\begin{proof}[Proof of Theorem \ref{thm:Pointlikes} \itref{bbb:PointlikesForRandLLevels} and \itref{ccc:PointlikesForRandLLevels}]
	We prove the result for \itref{bbb:PointlikesForRandLLevels}. The result for \itref{ccc:PointlikesForRandLLevels} is symmetric.
	From Lemma \ref{lem:inductionRandL} it follows that $\satR{m}M \subseteq \plR{m}M$ and from Lemma \ref{lem:inductionRandLother} it follows that $\pl{\tau}M \subseteq \satR{m}M$. By Proposition \ref{prp:RRcomparisons}, we have $N^{\XX}_{m,n} \in \mathbf{R}_{m+1}$ and thus $\plR{m}M \subseteq \pl{\tau}M$. In particular, $\plR{m}M = \pl{\tau}M = \satR{m}M$ which gives the desired result.
\end{proof}

To prove what is left of Theorem \ref{thm:Pointlikes}, we start by showing that every long enough $\XX_{m-1,n} \cup \YY_{m-1,n}$-set maps to an $\mathbf{RL}_m$-factor. The idea is analogous to the $\mathbf{R}_m$ and $\mathbf{L}_m$ cases.

\begin{lemma}\label{lem:ramseysappliedtwosided}
	Let $\mu: A^* \to M$ be a surjective morphism, and suppose that $M$ has a content morphism. Let $R$ be the Ramsey number of $2^M$, let $m \geq 2$ and let $n = (m-1+|A|)(R-1)$. If $U \subseteq A^*$ is $2(R-1)$-long and a $\XX_{m-1,n} \cup \YY_{m-1,n}$-set, then $\mu(U) \subseteq V$ where $V$ is an $\mathbf{RL}_m$-factor.
\end{lemma}

\begin{proof}
	Since $U$ is $2(R-1)$-long, we can factor 
	\begin{equation*}
		U = S_1 a_1 \dots S_{R-1} a_{R-1} Z b_1 T_1 \dots b_{R-1} T_{R-1}
	\end{equation*}
	where $\alp(S_i),\alp(T_i) \subsetneq \alp(S_ia_i) = \alp(b_iT_i) = \alp(U)$. Since each $a_i$ is reachable by a ranker $\mathsf{X}_{a_1 \ldots a_i}$ of depth at most $R-1$, it follows from Lemma \ref{lem:subwordsubrankerXX} that every $S_i$ is an $\XX_{m-1,n'}$-set where $n' = (m + |A| - 2)(R-1) = (m -1 + |\alp(S_i)|)(R-1)$. It follows from Lemma \ref{lem:inductionRandLother} that $\mu(S_i) \in \satR{m}M$. In particular, this implies that any product of $S_i$ and $a_j$ is in $\satR{m}M$. We get the symmetric result for products of $T_i$ and $b_j$; they are in $\satL{m}M$.

	By Lemma \ref{lem:ramseysapplied}, there exists $S_j a_j \dots S_{k} a_{k}$ such that $\mu(S_j a_j \dots S_{k} a_{k})$ is subidempotent. We set $S = \mu(S_1 a_1 \dots S_{j-1}a_{j-1})$ and note that $S \in \satR{m}M$ by the argument above. Let $E = \mu(S_i a_i \dots S_{i'} a_{i'})^{\theta}$ where $\theta = \omega_{2^M}$ is the idempotent power of $2^M$. Since $\mu(S_i a_i \dots S_{i'} a_{i'}) \in \satR{m}M$ and closed sets are closed under multiplication, we get $E \in \satR{m}M$. By subidempotency, it follows that $\mu(S_i a_i \dots S_{i'} a_{i'}) \subseteq E$.
	Symmetrically, we find $j'$, $k'$ such that $\mu(b_{j'} T_{j'} \dots b_{k'-1}T_{k'-1}) \subseteq F$ where $F \in \satL{m}M$ is idempotent and such that $T = \mu(b_{k'} T_{k'} \dots b_{R-1} T_{R-1}) \in \satL{m}M$. Finally, we set $W = S_{k+1} \dots S_R a_R Z b_1 T_1 \dots T_{j'-1}$. It follows that $U \subseteq SEWFT$ where $SEWFT = V$ is an $\mathbf{RL}_m$-factor.
\end{proof}

The desired result now follows using the factorizations obtained in Lemma \ref{lem:structurefactorization}. We give the result for the monoids $N^{\XY}_{m,n}$, but the proof generalizes directly to the other cases.

\begin{lemma}\label{lem:difficultdirectioncone}
	Let $M$ be a monoid with a content morphism $\alpha: M \to J_A$. Let $m \geq 2$ and let $\relmorsmall{\tau}{A^*}{M}{N^{\XY}_{m,n}}{\mu}{\nu}$ where $n = (m-1+3|A|)(R-1) + |A|$ depends on the Ramsey number $R$ of $2^M$. If $m$ is odd, then $\cone{\tau}{M} \subseteq \psatS{m}{M}$ and if $m$ is even, then $\cone{\tau}{M} \subseteq \psatP{m}{M}$.
\end{lemma}

\begin{proof}
	Let $U \subseteq A^*$ and $u \in U$ be such that $u \leq^{\XY}_{m,n} u'$ for all $u' \in U$ (in other words, $U$ is a $\XY_{m,n}$-set with minimal element $u$). We will prove $(\mu(u),\mu(U)) \in \psatS{m}M$ for odd $m$, and $(\mu(u),\mu(U)) \in \psatP{m}M$ for even $m$. Given $x \in N$, we can choose $u \in \nu^{-1}(x)$ such that $\nu^{-1}(\uparrow x)$ is a $\XY_{m,n}$-set with minimal element $u$, and thus this implies the desired result. We will use induction over $(m,|A|)$ ordered alphabetically. We consider the cases when $m$ is odd, with the even cases handled symmetrically.

	Let $m \geq 2$. We apply Lemma \ref{lem:structurefactorization} with $2(R-1)$ to obtain a factorization
	\begin{equation*}
		 U = U_1 b_1 V_1 a_1 \dots U_{k-1} b_{k-1} V_{k-1} a_{k-1} U_k
	\end{equation*}
	with the properties specified in the theorem. In particular, this gives a factorization
	\begin{equation*}
		u = u_1 b_1 v_1 a_1 \dots u_{k-1} b_{k-1} v_{k-1} a_{k-1} u_k.
	\end{equation*}	
	Every position marked by $a_i$ or $b_i$ in every $u'$ can be reached by a ranker of depth at most $(2(R-1)+1)|A\setminus\alp(V_i)|$. Since
	\begin{equation*}
		(m-1 + |A|)(R-1) = (m-1+3|A|)(R-1) + |A| - (2(R-1)+1)|A|
	\end{equation*}
	it follows that every $U_i$ is an $\XX_{m-1,(m-1+|A|)(R-1)} \cup \YY_{m-1,(m-1+|A|)(R-1)}$-set. By Lemma \ref{lem:ramseysappliedtwosided} we get an $\mathbf{RL}_m$-set $S_i$ such that $\mu(U_i) \subseteq S_i$.

	Let $A_i = \alp(V_i) \subsetneq A_i$, let $M_i = \mu(A_i^*)$ and let $\mu_i: A_i \to M_i$ be the restriction of $\mu$. For $m \geq 3$, we define
	\begin{align*}
		n_i & = (m-2+3|A_i|)(R_i-1) + |A_i|
		\\ & \leq (m-1+3|A|)(R-1) + |A| - 2(R-1)|A \setminus A_i|,
	\end{align*}
	where $R_i$ is the Ramsey number of $2^{M_i}$ and for $m = 2$, we define
	\begin{align*}
		n_i & = \ceil{R_i'/2} - 1
		\\ & \leq (m-1+3|A|)(R-1) + |A| - 2(R-1)|A \setminus A_i|,
	\end{align*}
	where $R_i'$ is the Ramsey number of $M_i$.
	We note that every $V_i$ is a $\XY_{m-1,n_i} \cup \YX_{m-1,n_i}$-set by Lemma \ref{lem:subwordsubranker}.
	If $|A| = 0$, then $\left( \mu(v_i), \mu(V_i) \right) = \left( \varepsilon, \left\{ \varepsilon \right\} \right)$ which is in any closed set. Otherwise, induction on $|A|$ gives $(t_i,T_i) \in \psatS{m-1}{M_i} \subseteq \psatS{m-1}M$ such that $\mu(v_i) = t_i$, $\mu(V_i) \subseteq T_i$. We define
	\begin{align*}
		 W & = S_1 \mu(b_1) T_1 \mu(a_1) \dots S_{k-1} \mu(b_{k-1}) T_{k-1} \mu(a_{k-1}) S_k,
		 \\ w & = s_1 \mu(b_1) t_1 \mu(a_1) \dots s_{k-1} \mu(b_{k-1}) t_{k-1} \mu(a_{k-1}) s_k.
	\end{align*}
	Suppose $a_j$ and $b_j$ conincide, and $a_{\ell}$ and $b_{\ell}$ coincide but that $a_{i}$ and $b_{i}$ does not coincide for any $j < i < \ell$. We claim that
	\begin{equation*}
		(\mu(u_jb_j),S_{j}\mu(b_j))(t_j,T_j) \cdots (t_{\ell-1},T_{\ell-1})(\mu(a_{\ell-1}u_\ell),\mu(a_{\ell-1})U_\ell)
	\end{equation*}
	is of the form required for $\psatS{m}M$ (we can use the same argument for $j = 1$ and $\ell = k$). Indeed, since $a_{i-1}, b_i \in \alpha(S_i) = \alp(U_i)$ and $S_i$ is an $\mathbf{RL}_m$ factor, it follows that $S_j\mu(b_j)$, $\mu(a_{i-1})S_i\mu(b_i)$ and $\mu(a_{\ell-1})S_{\ell}$ are $\mathbf{RL}_m$-factors. Since $\alp(v_i') \subseteq \alp(U_i) \cap \alp(U_{i+1})$ for all $v_i' \in V_i$, this gives a factorization of the desired form. 

	Since $(w,W)$ is a product of such factors and pairs $(\mu(a_i),\left\{ \mu(a_i) \right\})$, it follows that $(w,W) \in \psatS{m}M$. Since $\mu(u) = w$ $\mu(U) \subseteq W$ and closed sets are closed under subsets, this implies the desired result.
\end{proof}

\begin{proof}[Proof of Theorem \ref{thm:Pointlikes} \itref{ccc:ConelikesForSiAndPiLevels}, \itref{ccd:ConelikesForSiAndPiLevels}, \itref{ddd:ConelikesForSiAndPiLevels} and \itref{eee:ConelikesForSiAndPiLevels}]
	We prove the result for \itref{ddd:ConelikesForSiAndPiLevels}; the other results are analogous.

	From Lemma \ref{lem:satispointlikecone} it follows that $\psatS{m}M \subseteq \coneS{m}M$ and from Lemma \ref{lem:difficultdirectioncone} it follows that $\pl{\tau}M \subseteq \satR{m}M$ for $m$ odd. We get a similar result for $m$ even. In particular, $\plR{m}M = \pl{\tau}M = \satR{m}M$ which gives the desired result.
\end{proof}

\section{Conclusion and Outlook}

We considered conelikes, an algebraic counterpart to the covering problem. In particular, solving the conelike problem yields solutions to the separation problem for positive varieties. For full varieties, this problem coincides with the pointlike problem. We provided solutions to the conelike (resp.\ pointlike) problem for all levels of the Trotter-Weil and the quantifier alternation hierarchy (Theorem \ref{thm:Pointlikes}). This was done by providing computable subsets of $M \times 2^M$ (resp.\ $2^M$) and showing that these coincided with the conelikes (resp.\ pointlikes).

Furthermore, we considered comparisons of rankers. We showed that any set of ranker comparisons closed under subwords gives rise to a stable preorder, and thus a monoid. The quantifier alternation hierarchy has previously been given a characterization in terms of ranker comparisons. We extended this to a characterization using ranker comparisons for the corners of the Trotter-Weil hierarchy. Apart from giving a unifying picture of the two hierarchies as a ranker comparison hierarchy (Figure \ref{fig:hierarchies}), this also served as a tool for the result on pointlikes. Having a unified formalism made moving up in the hierarchy much more uniform.
We also used ranker comparisons to find separators for the conelike problem, i.e.\ relational morphisms such that a set is conelike with respect to that relational morphism if and only if it is conelike with respect to the variety.

Given a separator, one can try all possible quotients in order to find an \emph{optimal separator}, i.e., the smallest monoid acting as a separator. However, the monoids provided here are doubly exponential in the size of $M$ (exponential in the size $n$ of the  rankers which in turn is exponential in $M$), making such an approach computationally hardly feasable. Thus, in future work, it would be interesting to get a better understanding of optimal separators.

\bibliographystyle{abbrv}
\bibliography{library}

\end{document}